\documentclass[preprint,showkeys,lengthcheck,nofootinbib,twocolumn,notitlepage,floatfix,superscriptaddress]{revtex4-1}
\usepackage{graphics,graphicx}
\usepackage{amsmath, amssymb}
\usepackage{multirow}
\usepackage{color}
\usepackage{verbatim}
\usepackage{hyperref}
\usepackage[normalem]{ulem}  
\usepackage{ulem}
\usepackage{color}
\usepackage{cancel}
\usepackage{mathtools}
\usepackage{epstopdf}

\renewcommand\sout{\bgroup\color{blue} \ULdepth=-.5ex \ULset}
\def\slashchar#1{\setbox0=\hbox{$#1$}  
\dimen0=\wd0     
\setbox1=\hbox{/} \dimen1=\wd1  
\ifdim\dimen0>\dimen1   
\rlap{\hbox to \dimen0{\hfil/\hfil}} 
#1     
\else     
\rlap{\hbox to \dimen1{\hfil$#1$\hfil}} 
/      
\fi}

\newcommand{\dd}{\mathrm{d}}
\newcommand{\pp}{\partial}

\begin{document}

\title{Fluctuations near the liquid-gas and chiral phase transitions in hadronic matter}

\date{\today}
\author{Micha\l{} Marczenko}
\email{michal.marczenko@uwr.edu.pl}
\address{Incubator of Scientific Excellence - Centre for Simulations of Superdense Fluids, University of Wroc\l{}aw, plac Maksa Borna 9, PL-50204 Wroc\l{}aw, Poland}
\author{Krzysztof Redlich}
\address{Institute of Theoretical Physics, University of Wroc\l{}aw, plac Maksa Borna 9, PL-50204 Wroc\l{}aw, Poland}
\address{Polish Academy of Sciences PAN, Podwale 75, 
PL-50449 Wroc\l{}aw, Poland}
\author{Chihiro Sasaki}
\address{Institute of Theoretical Physics, University of Wroc\l{}aw, plac Maksa Borna 9, PL-50204 Wroc\l{}aw, Poland}
\address{International Institute for Sustainability with Knotted Chiral Meta Matter (SKCM$^2$), Hiroshima University, Higashi-Hiroshima, Hiroshima 739-8511, Japan}

\begin{abstract}
We investigate the fluctuations of the net-baryon number density in dense hadronic matter. Chiral dynamics is modeled via the parity doublet Lagrangian, and the mean-field approximation is employed to account for chiral criticality. We focus on the qualitative properties and systematics of the second-order susceptibility of the net-baryon number density for individual positive- and negative-parity nucleons whose masses become degenerate at the chiral restoration. It is shown that the second-order susceptibility of the positive-parity state can become negative when the chiral symmetry is restored, as a natural consequence of the unique relationship of the mass to the order parameter. Moreover, we find that such negative fluctuations are indicative of approaching the critical point on the chiral phase boundary. Our results may have consequences for the interpretation of the experimental data on net-proton fluctuations in heavy-ion collisions.
\end{abstract}
\maketitle

\section{Introduction}

Understanding the thermodynamic properties of strongly interacting matter, described by quantum chromodynamics (QCD), is a formidable task. One of the challenges in modern high-energy physics is to determine the QCD phase diagram with an anticipated critical point at finite net baryon density. At vanishing net-baryon number density, a reliable description has been provided through the first-principle lattice QCD (LQCD) calculations, which shows that the equation of state (EoS) exhibits a smooth crossover from confined hadronic matter to deconfined quark-gluon plasma~\cite{Bazavov:2014pvz, Borsanyi:2018grb, Bazavov:2017dus, Bazavov:2020bjn, Bazavov:2020bjn}. This transition is linked to the simultaneous onset of chiral symmetry restoration and quark deconfinement~\cite{Aoki:2006we, Bazavov:2018mes}. However, the LQCD approach remains insufficient to determine the nature of the EoS at finite density owing to the sign problem,
and the existence of any QCD critical point(s) is unresolved.

Observables associated with fluctuations and correlations of conserved charges are promising for the search of the chiral-critical behavior at the QCD phase boundary~\cite{Stephanov:1999zu, Asakawa:2000wh, Hatta:2003wn, Friman:2011pf}, and chemical freeze-out of produced hadrons in heavy-ion collisions (HIC)~\cite{Bazavov:2012vg, Borsanyi:2014ewa, Karsch:2010ck, Braun-Munzinger:2014lba, Vovchenko:2020tsr, Braun-Munzinger:2020jbk}. In particular, fluctuations have been proposed to probe the QCD critical point, as well as the remnants of the $O(4)$ criticality at vanishing and finite net-baryon densities~\cite{Friman:2011pf, Stephanov:2011pb, Karsch:2019mbv, Braun-Munzinger:2020jbk, Braun-Munzinger:2016yjz}. 
The search for a critical point has been extensively conducted in HIC within the beam energy scan (BES) programs at the Relativistic Heavy Ion Collider (RHIC) at BNL~\cite{STAR2010}  and the Super Proton Synchrotron (SPS) at CERN~\cite{Mackowiak-Pawlowska:2020glz}. However, no conclusive evidence has been observed so far for a critical point.

At small net-baryon number density, the QCD thermodynamics in the confined phase is well-described by the hadron resonance gas (HRG) model~\cite{BraunMunzinger:2003zd, Andronic:2017pug}. The HRG model explains satisfactorily the LQCD data below the crossover to the quark-gluon plasma, as well as various hadron yields in HIC~\cite{Andronic:2017pug}. Several extensions of the HRG model have been proposed to quantify the LQCD EoS and various fluctuation observables up to near-chiral crossover. They account for consistent implementation of hadronic interactions within the S-matrix approach~\cite{Venugopalan:1992hy, Broniowski:2015oha, Friman:2015zua, Huovinen:2016xxq, Lo:2017lym}, a more complete implementation or a continuously growing exponential mass spectrum and/or possible repulsive interactions among constituents~\cite{Majumder:2010ik, Andronic:2012ut, Albright:2014gva, Vovchenko:2014pka, Lo:2015cca, ManLo:2016pgd, Andronic:2020iyg}. Nevertheless, it is challenging to identify the role of different in-medium effects and hadronic interactions on the properties of higher-order fluctuations of conserved charges. Recently, it was argued that deviations of the LQCD data on higher-order fluctuations of net-baryon number density from the HRG baseline in the near vicinity of the chiral transition can be attributed to repulsive interactions~\cite{Vovchenko:2016rkn}. However, an adequate description of the higher-order susceptibilities of the net-baryon density in the chiral crossover requires a more refined framework that accounts for a self-consistent treatment of the chiral in-medium effects and repulsive interactions~\cite{Marczenko:2021icv}.

How does the chiral symmetry restoration become manifest in the baryon masses?
The LQCD results~\citep{Aarts:2015mma, Aarts:2017rrl, Aarts:2018glk} exhibit a clear emergence of the parity doubling structure for the low-lying baryons around the chiral crossover. The masses of the positive-parity ground states are found to be rather weakly temperature-dependent, while the masses of negative-parity states drop substantially when approaching the chiral crossover temperature $T_c$. The parity doublet states become almost degenerate with a finite mass in the vicinity of the chiral crossover. Even though these LQCD results are still not obtained in the physical limit, the observed behavior of parity partners is likely an imprint of the chiral symmetry restoration in the baryonic sector of QCD. Such properties of the chiral partners can be described in the framework of the parity doublet model~\citep{Detar:1988kn, Jido:1999hd, Jido:2001nt}.
The model has been applied to the vacuum phenomenology of QCD, hot and dense hadronic matter, as well as neutron stars~\citep{Dexheimer:2007tn, Gallas:2009qp, Paeng:2011hy, Sasaki:2011ff, Gallas:2011qp, Zschiesche:2006zj, Benic:2015pia, Marczenko:2017huu, Marczenko:2018jui, Marczenko:2019trv, Marczenko:2020wlc, Marczenko:2020jma, Marczenko:2021uaj, Marczenko:2022hyt, Mukherjee:2017jzi, Mukherjee:2016nhb, Dexheimer:2012eu, Steinheimer:2011ea, Weyrich:2015hha, Sasaki:2010bp, Yamazaki:2018stk, Yamazaki:2019tuo, Ishikawa:2018yey, Steinheimer:2010ib, Giacosa:2011qd, Motohiro:2015taa, Minamikawa:2020jfj}.

In this work, we analyze the qualitative properties and systematics of the fluctuations of conserved charges in the context of the parity doublet model, which incorporates the chiral symmetry restoration and repulsive interactions via the exchange of the scalar and vector mesons, respectively. To account for critical behaviors, the mean-field approximation is employed, which captures the same characteristics as those of the $O(4)$ criticality, albeit with different critical exponents.
We study the properties of the second-order susceptibility of the net-baryon number density for positive- and negative-parity nucleons, individually. Their qualitative behavior is examined near the chiral, as well as the nuclear liquid-gas phase transitions.

This paper is organized as follows. In Sec.~\ref{sec:pd_model}, we introduce the parity doublet model. In Sec.~\ref{sec:susceptibility}, we discuss the structure of the susceptibilities of the net-baryon number density. In Sec.~\ref{sec:results}, we present our results and clarify the role of the nucleon parity doublet near the two phase transitions. Finally, Sec.~\ref{sec:summary} is devoted to summary and conclusions.

\section{Parity doublet model}
\label{sec:pd_model}

In the conventional Gell-Mann--Levy model of mesons and nucleons~\cite{GellMann:1960np}, the nucleon mass is entirely generated by the non-vanishing expectation value of the sigma field. Thus, the nucleon inevitably becomes massless when the chiral symmetry gets restored. This is led by the particular chirality assignment to the nucleon parity doublers, where the nucleons are assumed to be transformed in the same way as the quarks are under chiral rotations.

More general allocation of the left- and right-handed chiralities to the nucleons, the mirror assignment, was proposed in~\cite{Detar:1988kn}. This allows an explicit mass term for the nucleons, and consequently, the nucleons stay massive at the chiral restoration point. For more details, see Refs.~\cite{Detar:1988kn,Jido:1999hd,Jido:2001nt}.
	
\begin{table*}[t!]\begin{center}\begin{tabular}{|c|c|c|c|c|c|c|c|c|c|c|}
\hline
$m_0~$[MeV] & $m_+~$[MeV] & $m_-~$[MeV] & $m_\pi~$[MeV] & $f_\pi~$[MeV] & $m_\omega~$[MeV] & $\lambda_4$ & $\lambda_6f_\pi^2$ & $g_\omega$ & $g_1$ & $g_2$ \\ \hline\hline
750 & 939   & 1500  & 140     & 93      & 783        &   28.43	 &   11.10       &    6.45    & 13.36 & 7.32 \\ \hline
\end{tabular}\end{center}
\caption{Physical inputs in matter-free space and the model parameters used in this work. See Sec.~\ref{sec:pd_model} for details.}
\label{tab:vacuum_params}
\end{table*}

In the mirror assignment, under \mbox{$SU(2)_L \times SU(2)_R$} rotation, two chiral fields $\psi_1$ and $\psi_2$ are transformed as follows:
\begin{equation}\label{eq:mirror_assignment}
\begin{split}
    \psi_{1L} \rightarrow L\psi_{1L}, \;\;\;\; \psi_{1R} \rightarrow R\psi_{1R}\textrm, \\
    \psi_{2L} \rightarrow R\psi_{2L}, \;\;\;\; \psi_{2R} \rightarrow L\psi_{2R}\textrm,
\end{split}
\end{equation}
where $\psi_i = \psi_{iL} + \psi_{iR}$, $L \in SU(2)_L$ and $R \in SU(2)_R$. In this work, we consider a system with $N_f = 2$, hence, relevant for this study are the lowest nucleons and their chiral partners. The hadronic degrees of freedom are coupled to the chiral fields $(\sigma,~\pi)$, and the iso-singlet vector field $\omega_\mu$. The nucleon part of the Lagrangian in the mirror model reads
\begin{equation}\label{eq:doublet_lagrangian}
\begin{split}
    \mathcal{L}_N &= i\bar\psi_1\slashchar\partial\psi_1 + i\bar\psi_2\slashchar\partial\psi_2 + m_0\left(  \bar\psi_1\gamma_5\psi_2 - \bar\psi_2\gamma_5\psi_1 \right) \\
    &+ g_1\bar\psi_1 \left( \sigma + i\gamma_5 \boldsymbol\tau \cdot \boldsymbol\pi \right)\psi_1 + g_2\bar\psi_2 \left( \sigma - i\gamma_5 \boldsymbol\tau \cdot \boldsymbol\pi \right)\psi_2 \\
    &-g_\omega\bar\psi_1\slashchar\omega\psi_1 - g_\omega\bar\psi_2\slashchar\omega\psi_2 \textrm,
\end{split}
\end{equation}
where $g_1$, $g_2$, and $g_\omega$ are the baryon-to-meson coupling constants and $m_0$ is a mass parameter.

The mesonic part of the Lagrangian reads
\begin{equation}
\begin{split}
    \mathcal{L}_M = \frac{1}{2} \left( \partial_\mu \sigma\right)^2 + \frac{1}{2} \left(\partial_\mu \boldsymbol\pi \right)^2 - \frac{1}{4} \left( \omega_{\mu\nu}\right)^2-V_\sigma - V_\omega \textrm,
\end{split}
\end{equation}
where $\omega_{\mu\nu} = \partial_\mu\omega_\nu - \partial_\nu\omega_\mu$ is the field-strength tensor of the vector field, and the potentials read
\begin{subequations}\label{eq:potentials_parity_doublet}
\begin{align}
    V_\sigma &= -\frac{\lambda_2}{2}\Sigma + \frac{\lambda_4}{4}\Sigma^2 - \frac{\lambda_6}{6}\Sigma^3- \epsilon\sigma \textrm,\label{eq:potentials_sigma}\\
    V_\omega &= -\frac{m_\omega^2 }{2}\omega_\mu\omega^\mu\textrm.
\end{align}
\end{subequations}
where $\Sigma = \sigma^2 + \boldsymbol\pi^2$, $\lambda_2 = \lambda_4f_\pi^2 - \lambda_6f_\pi^4 - m_\pi^2$, and $\epsilon = m_\pi^2 f_\pi$. $m_\pi$ and $m_\omega$ are the $\pi$ and $\omega$ meson masses, respectively, and $f_\pi$ is the pion decay constant. Note that the chiral symmetry is explicitly broken by the linear term in $\sigma$ in Eq.~\eqref{eq:potentials_sigma}.

\begin{figure}
    \centering
    \includegraphics[width=\linewidth]{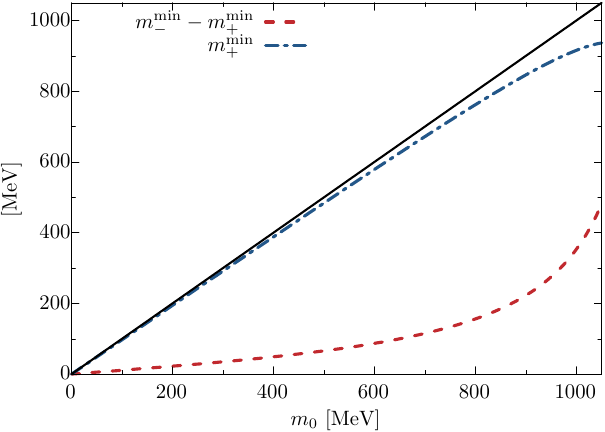}
    \caption{The minimum mass $m_+^{\rm min}$ and the order parameter $\delta^{\rm min}_\sigma = m_-^{\rm min} - m_+^{\rm min}$ as a function of the chirally invariant mass $m_0$. The black, solid line shows $m_+^{\rm min} = m_0$.}
    \label{fig:sigma0}
\end{figure}

The full Lagrangian of the parity doublet model is given by
\begin{equation}
    \mathcal L = \mathcal L_N + \mathcal L_M\textrm.
\end{equation}

In the diagonal basis, the masses of the positive- and negative-parity baryonic chiral partners, $N_\pm$, are given by	
\begin{equation}\label{eq:doublet_masses}
	m_\pm = \frac{1}{2}\left(\sqrt{\alpha^2\sigma^2 + 4m_0^2} \mp \beta\sigma\right) \textrm,
\end{equation}
where $\alpha = g_1+g_2$, $\beta=g_1-g_2$. From Eq.~(\ref{eq:doublet_masses}), it is clear that, in contrast to the naive assignment under chiral symmetry, the chiral symmetry breaking generates only the splitting between the two masses. When the symmetry is restored, the masses become degenerate, $m_\pm(\sigma=0) = m_0$.

To investigate the properties of strongly-interacting matter, we adopt the mean-field approximation~\cite{Serot:1984ey}. Rotational invariance requires that the spatial component of the $\omega_\mu$ field vanishes, namely $\langle \boldsymbol \omega \rangle = 0$\footnote{Since $\omega_0$ is the only non-zero component in the mean-field approximation, we simply denote it by $\omega_0 \equiv\omega$.}. Parity conservation on the other hand dictates $\langle \boldsymbol \pi \rangle = 0$. The mean-field thermodynamic potential of the parity doublet model reads
\begin{equation}\label{eq:thermo_potential}
\Omega = \Omega_+ + \Omega_- + V_\sigma + V_\omega \textrm,
\end{equation}
with
\begin{equation}\label{eq:kinetic_thermo}
\Omega_\pm = \gamma_\pm \int\frac{\dd^3 p}{(2\pi)^3}\; T \left[ \ln\left(1 - f_\pm\right) + \ln\left(1 - \bar f_\pm\right) \right]\textrm,
\end{equation}
where $\gamma_\pm = 2\times 2$ denotes the spin-isospin degeneracy factor for both parity partners, and $f_\pm$  $(\bar f_\pm)$ is the particle (antiparticle) Fermi-Dirac distribution function,
\begin{equation}\label{eq:fermi_dist_nucleon}
\begin{split}
f_\pm = \frac{1}{1+ e^{\left(E_\pm - \mu^\ast\right)/T}} \textrm,\\
\bar f_\pm = \frac{1}{1+ e^{\left(E_\pm + \mu^\ast\right)/T}}\textrm, \\
\end{split}
\end{equation}
where $T$ is the temperature, the dispersion relation $E_\pm = \sqrt{\boldsymbol p^2 + m_\pm^2}$,  and the effective baryon chemical potential $\mu^\ast = \mu_B - g_\omega \omega$.

\begin{figure*}[ht!]
    \centering
    \includegraphics[width=.49\linewidth]{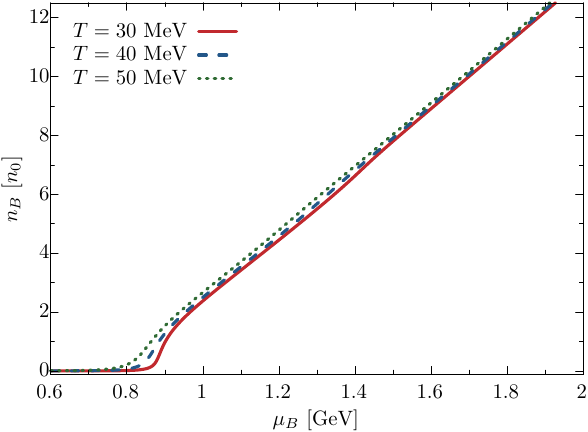}\;
    \includegraphics[width=.495\linewidth]{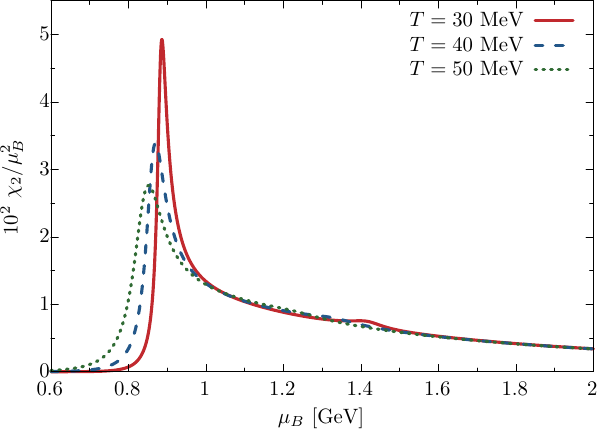}
    \caption{Net-baryon density (left panel) and its baryon-chemical-potential-normalized second-order susceptibility (right panel) plotted for fixed temperature.}
    \label{fig:nb_x2_full}
\end{figure*}

In-medium profiles of the mean fields are obtained by extremizing the thermodynamic potential in Eq.~\eqref{eq:thermo_potential}, leading to the following gap equations:
\begin{equation}
\begin{split}\label{eq:gap_eqs}
0=\frac{\partial \Omega}{\partial \sigma} &= \frac{\partial V_\sigma}{\partial \sigma} + s_+ \frac{\partial m_+}{\partial \sigma} + s_- \frac{\partial m_-}{\partial \sigma} \textrm,\\
0=\frac{\partial \Omega}{\partial \omega} &= \frac{\partial V_\omega}{\partial \omega} + g_\omega \left(n_B^+ + n_B^-\right) \textrm,
\end{split}
\end{equation}
where the scalar and vector densities are
\begin{equation}
s_\pm = \gamma_\pm \int \frac{\mathrm{d}^3 p}{\left(2\pi\right)^3} \frac{m_\pm}{E_\pm}\left(f_\pm + \bar f_\pm\right)
\end{equation}
and
\begin{equation}
n^\pm_B = \gamma_\pm \int \frac{\mathrm{d}^3 p}{\left(2\pi\right)^3}\left(f_\pm - \bar f_\pm\right) \textrm,
\end{equation}
respectively.

In the grand canonical ensemble, the thermodynamic pressure reads
\begin{equation}\label{eq:pressure}
P= -\Omega + \Omega_0\textrm,
\end{equation}
where $\Omega_0$ is the value of the thermodynamic potential in the vacuum, and the net-baryon number density can be calculated as follows:
\begin{equation}\label{eq:nb}
n_B = \frac{\pp P(T, \mu_B)}{\pp \mu_B} = n_B^+ + n_B^-\textrm,
\end{equation}

The positive-parity state, $N_+$, corresponds to the nucleon $N(938)$. Its negative parity partner is identified with $N(1535)$. Their vacuum masses are shown in Table~\ref{tab:vacuum_params}. The value of the parameter $m_0$ has to be chosen so that a chiral crossover is realized at finite temperature and vanishing chemical potential. The model predicts the chiral symmetry restoration to be a crossover for $m_0\gtrsim 700~$MeV. Following the previous studies of the \mbox{parity-doublet-based} models~\cite{Dexheimer:2007tn,Gallas:2009qp,Paeng:2011hy, Sasaki:2011ff, Gallas:2011qp, Zschiesche:2006zj, Benic:2015pia, Marczenko:2017huu, Marczenko:2018jui, Marczenko:2019trv, Marczenko:2020wlc, Marczenko:2020jma, Motornenko:2019arp, Mukherjee:2017jzi, Mukherjee:2016nhb, Dexheimer:2012eu, Steinheimer:2011ea, Weyrich:2015hha, Sasaki:2010bp, Yamazaki:2018stk,Yamazaki:2019tuo, Ishikawa:2018yey, Steinheimer:2010ib, Giacosa:2011qd,Motohiro:2015taa,Minamikawa:2020jfj}, as well as recent lattice QCD results~\cite{Aarts:2017rrl, Aarts:2018glk,Aarts:2015mma}, we choose a rather large value, $m_0=750$~MeV. We note, however, that the results presented in this work qualitatively do not depend on the choice of $m_0$, as long as the chiral crossover appears at $\mu_B=0$. The parameters $g_1$ and $g_2$ are determined by the aforementioned vacuum nucleon masses and the chirally invariant mass $m_0$ via Eq.~\eqref{eq:doublet_masses}. The parameters $g_\omega$, $\lambda_4$ and $\lambda_6$ are fixed by the properties of the nuclear ground state at zero temperature, i.e., the saturation density, binding energy, and compressibility parameter at $\mu_B=923~$MeV. The constraints are as follows:
\begin{subequations}
\begin{align}
n_B &= 0.16~\textrm{fm}^{-3}\textrm,\\
E/A - m_+ &= -16~\textrm{MeV}\textrm,\\
K = 9n^2_B\frac{\pp^2\left( E/A \right)}{\pp n_B^2} &= 240~\textrm{MeV}\textrm.\label{eq:compres}
\end{align}
\end{subequations}
We note that the six-point scalar interaction term in Eq.~\eqref{eq:potentials_sigma} is essential to reproduce the empirical value of the compressibility in Eq.~\eqref{eq:compres}~\citep{Motohiro:2015taa}. A compilation of the parameters used in this paper is found in Table~\ref{tab:vacuum_params}. For this set of parameters, we obtain the pseudo-critical temperature of the chiral crossover at vanishing chemical potential, $T_c=209~$MeV. At low temperature, the model predicts sequential first-order liquid-gas and chiral phase transitions with critical points located at $T_{\rm lg} = 16~$MeV, $\mu_B=909~$MeV, ($n_B=0.053~\textrm {fm}^{-3}=0.33n_0$) and $T_{\rm ch} = 7~$MeV, $\mu_B=1526~$MeV ($n_B=1.25~\textrm{fm}^{-3}=7.82n_0$), respectively.

In the next section, we discuss the general structure of the second-order susceptibilities of the net-baryon number density for positive- and negative-parity chiral partners to quantify their roles near the second-order phase transition at finite density.

\begin{figure*}[ht!]
    \centering
    \includegraphics[width=.49\linewidth]{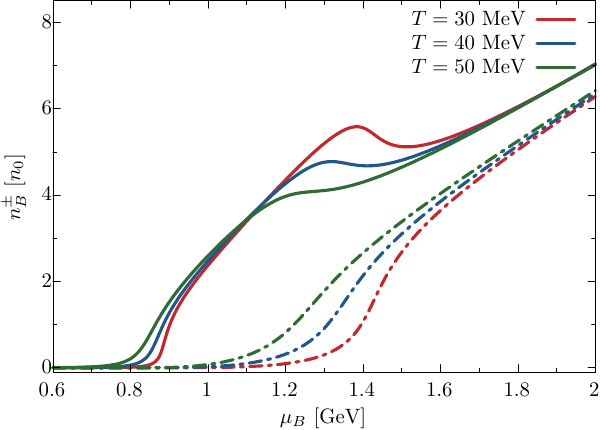}\;
    \includegraphics[width=.49\linewidth]{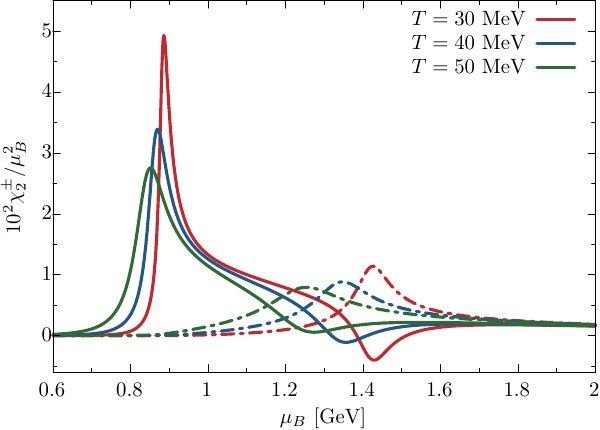}
    \caption{Net-baryon density (left panel) and its baryon-chemical-potential-normalized second-order susceptibility (right panel) for positive- and negative-parity chiral partners separately plotted for fixed temperature. Solid and dash-dotted lines correspond to positive- and negative-parity chiral partners, respectively.}
    \label{fig:nb_x2}
\end{figure*}

\section{Net-baryon number susceptibility}\label{sec:susceptibility}

The main objective of the present studies is to analyze and delineate the contribution of nucleon parity doubler to fluctuations of the net-baryon number density at finite temperature and baryon chemical potential. In general, the fluctuations of conserved charges reveal more information about the matter composition than the equation of state and can be used as probes of a phase boundary. The critical properties of chiral models, within the functional renormalization group (FRG) approach~\cite{Wetterich:1992yh, Morris:1993qb, Ellwanger:1993mw, Berges:2000ew}, are governed by the same universality classes as in QCD, i.e., the chiral transition belongs to $O(4)$ universality class, which, at large values of the baryon chemical potential, may develop a $Z(2)$ critical point, followed by the first-order phase transition~\cite{Asakawa:1989bq, Halasz:1998qr, Berges:1998rc}. This criticality is naturally encoded in quark-based models~\cite{Skokov:2010wb, Skokov:2010uh, Schaefer:2006ds, Friman:2011pf, Almasi:2017bhq}, as well as the hadronic parity doublet model. We note that the mean-field treatment yields different critical exponents, albeit preserving the structure of the phase diagram.

In the grand canonical ensemble, the generalized susceptibilities of the net-baryon number, $\chi_n$, are defined as derivatives with respect to the baryon chemical potential,
\begin{equation}\label{eq:fluct_def}
    \chi_n\left(T, \mu_B\right) = \frac{\partial^{n-1} n_B\left(T, \mu_B\right)}{\partial \mu_B^{n-1}} \Bigg|_T = \chi^+_n + \chi_n^-\textrm.
\end{equation}

The net-baryon number density, as well as any other thermodynamic quantity, contains explicit dependence on the mean fields. In this work, we consider the isospin-symmetric matter, therefore relevant are the scalar $\sigma$ and vector $\omega$ mean fields, i.e. \mbox{$n_B = n_B\left(T, \mu_B, \sigma(T, \mu_B), \omega(T, \mu_B)\right)$}. Consequently, the second-order susceptibility, $\chi_2^\pm$, can be written explicitly as
\begin{equation}\label{eq:chi_2_i}
    \chi^\pm_2 = \frac{\pp n_B^\pm}{\pp \mu_B} + \frac{\pp n_\pm}{\pp m_\pm}\frac{\pp m_\pm}{\pp\sigma}\frac{\pp \sigma}{\pp \mu_B} + \frac{\pp n_\pm}{\pp \omega}\frac{\pp \omega}{\pp \mu_B}
\end{equation}
The middle term in Eq.~\eqref{eq:chi_2_i} is the chiral-critical mode:
\begin{equation}\label{eq:chi_2_crit}
    \chi_2^{\pm, \;\rm crit} \sim \frac{\pp n_\pm}{\pp m_\pm}\frac{\pp m_\pm}{\pp\sigma}\frac{\pp \sigma}{\pp \mu_B}
\end{equation}
The derivative $\pp m_\pm / \pp \sigma$ is readily calculated from Eq.~\eqref{eq:doublet_masses}, namely
\begin{equation}
    \frac{\pp m_\pm}{\pp\sigma} = \frac{1}{2}\left(\frac{\alpha^2\sigma}{\sqrt{\alpha^2\sigma^2 + 4m_0^2}}\mp \beta \right) \textrm.
\end{equation}
Note that for the positive-parity state, a minimum value of the mass, $m_+^{\rm min}$, exists at
\begin{equation}
\sigma_{\rm min} = \frac{2\beta m_0}{\alpha \sqrt{\alpha^2-\beta^2}} \textrm,
\end{equation}
while the mass of the negative-parity state monotonically decreases with $\sigma$ as the chiral symmetry gets restored. We also note that $\sigma_{\rm min} > 0$, therefore the chiral-critical mode for positive-parity state $\chi_2^{+, \;\rm crit}$ becomes negative when $\sigma < \sigma_{\rm min}$ is realized. 

In Fig.~\ref{fig:sigma0}, we show the threshold value of the order parameter, $\delta_\sigma \equiv m_- - m_+ = \beta \sigma$, at the minimum $\sigma_{\rm min}$ as a function of the chirally invariant mass. In general, it grows with $m_0$, which means that $\chi_2^{+,\rm crit}$ becomes negative when the chiral symmetry is more readily broken. For instance, for $m_0 = m^{\rm vac}_+ = 939~$MeV, $\delta_\sigma^{\rm min} = 262~$MeV and for $m_0=750~$MeV, $\delta_\sigma^{\rm min} = 133~$MeV. We note that the value of $m_+^{\rm min}$ is always below but close to $m_0$, regardless of the choice of the chirally invariant mass. Thus, the mass of the positive-parity nucleon attains its minimal value near the chiral restoration.

In general, the chiral-critical mode itself becomes negative at any temperature and baryon chemical potential when $\sigma < \sigma_{\rm min}$. Only in the vicinity of the critical point, it becomes substantially large as compared to the other terms in Eq.~\eqref{eq:chi_2_i} and divergent via the term $\partial \sigma / \partial \mu_B$. Moreover, at densities close to the liquid-gas phase transition, chiral symmetry is still to a large extent broken (i.e., $\sigma \approx \sigma_{\rm vac} > \sigma_{\rm min}$) and the fluctuations of the positive-parity state are expected to be positive as the critical point of the liquid-gas phase transition is approached. Therefore, $\chi_2^+$ becomes negative in the vicinity of the critical region of the chiral phase transition, where the term $\chi_2^{\pm,\;\rm crit}$ becomes dominant and changes sign. We note that for $m_0=0$, the order parameter $\delta_\sigma^{\rm min}=0$, which means that the $\chi_2^+$ is positive-defined. We also remark that the minimal value of $m_+$ is very close to the $m_0$ (see Fig.~\ref{fig:sigma0}) and depends only mildly on the value of $m_0$.

We emphasize that, in general, if the minimum of $m_+(\sigma)$ is reached at values of $T$ and $\mu_B$ which are close to the phase boundary, the properties discussed above are expected to appear independently of the position of the critical point on the phase diagram. Although the dependence of $m_+$ on $\sigma$ is not universal and model dependent, we stress that the calculations with the functional renormalization group techniques preserve the same in-medium behavior~\cite{Tripolt:2021jtp}. At present, however, the only reliable answer can be obtained from the first-principle Lattice QCD calculations.

In the following, we quantify the contributions of positive- and negative-parity chiral partners to the second-order susceptibility of the net-baryon number density in the vicinity of the nuclear liquid-gas and chiral phase transitions to identify the importance of the chiral-criticality.

\section{Results}\label{sec:results}

\begin{figure*}
    \centering
    \includegraphics[width=.49\linewidth]{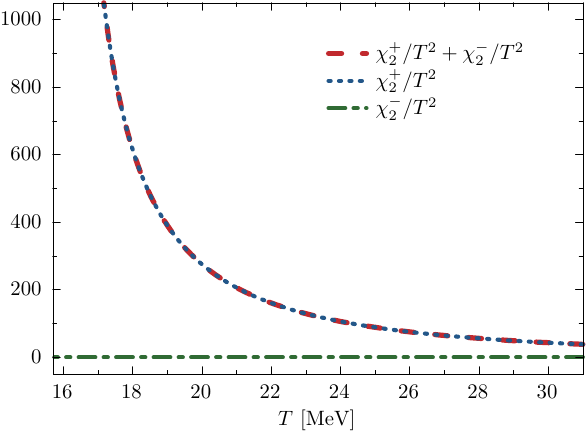}\;
    \includegraphics[width=.49\linewidth]{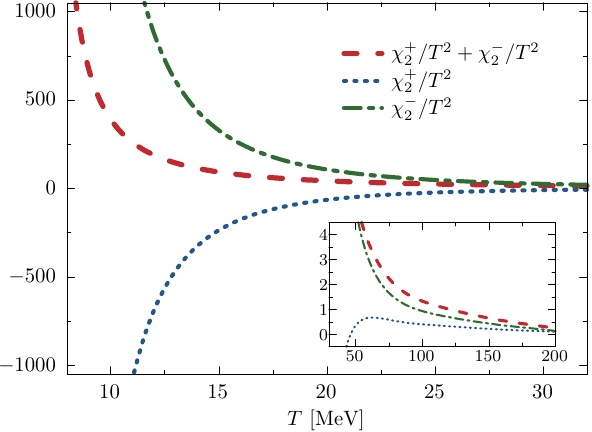}
    \caption{The temperature dependence of the temperature-normalized second-order susceptibility of the net-baryon number density along the crossover liquid-gas (left panel) and chiral (right panel) transition lines. In the right panel, the inlet figure shows the same quantities for higher temperatures.}
    \label{fig:chi2}
\end{figure*}

In the left panel of Fig.~\ref{fig:nb_x2_full}, we show the net-baryon number density at three different temperatures as a function of the baryon chemical potential. In general, $n_B$ features a rapid increase at small values of $\mu_B \approx 0.9~$GeV, which is a remnant of the liquid-gas phase transition at smaller temperatures. In the right panel of Fig.~\ref{fig:nb_x2_full}, we show the corresponding second-order susceptibility, $\chi_2$, normalized by the baryon chemical potential. It features a peak below $\mu_B\sim 1$~GeV, which corresponds to the rapid increase of $n_B$. At higher baryon chemical potential, $\chi_2$ shows only a mild peak around $\mu_B \approx 1.3 - 1.4~$GeV, which is a remnant of the chiral phase transition. Note that $\chi_2$ stays positive at all values of the baryon chemical potential.

More structure is revealed when contributions from positive- and negative-parity chiral partners are considered separately. This is shown in Fig.~\ref{fig:nb_x2}. In the left panel, we show the net densities, $n_\pm$. The net density $n_+$ features a rapid increase at small values of $\mu_B$, which signals a population of the positive-parity state. Likewise, a similar increase is seen in the net density $n_-$ at higher $\mu_B$. At $T=50~$MeV, the population of the negative-parity state additionally softens the EoS, which is also reflected in a slower increase of $n_+$. On the other hand, at $T=30$ and $40~$MeV, $n_+$ develops a local maximum followed by a local minimum. This is connected with a more rapid increase of $n_-$ at smaller temperatures. At high baryon chemical potential, chiral partners become equally populated due to chiral symmetry restoration. In the right panel of Fig.~\ref{fig:nb_x2}, we show the corresponding second-order susceptibilities of the net-baryon number density. The susceptibility $\chi_+$ features a peak below $\mu_B\sim 1$~GeV, which is a remnant of the liquid-gas phase transition. A similar peak is seen for $\chi_-$ at higher baryon chemical potential, which can be interpreted as a remnant of the chiral phase transition. Notably, at a chemical potential where $\chi_-$ features a peak, $\chi_+$ features a minimum. The value of $\chi_+$ at the minimum decreases with lowering temperature and eventually becomes negative. Therefore, the second-order susceptibility of the net-baryon number of positive-parity state behaves differently in the vicinity of liquid-gas and chiral phase transitions.

The fluctuations of the positive-parity nucleon manifest the onset of liquid-gas and chiral phase transitions in a different manner. To quantify the differences, we calculate the fluctuations as functions of temperature along the trajectories obtained by tracing the remnants of these two transitions, i.e., the corresponding peaks in $\chi_2^\pm$. The temperature dependence of $\chi_2^\pm$ along the remnant of the liquid-gas phase transition is shown in the left panel of Fig.~\ref{fig:chi2}. The susceptibility $\chi_2^+$ increases toward the critical point of the liquid-gas phase transition, located around $T=16~$MeV. On the other hand, $\chi_2^-$ stays roughly constant around zero, due to thermal suppression of the negative-parity state. Therefore, the fluctuations around the critical point of the liquid-gas phase transition are entirely driven by the fluctuations of the positive-parity state. In the right panel of Fig.~\ref{fig:chi2}, we show $\chi^\pm_2$ along the chiral crossover line. The entire $\chi_2$ diverges at the critical point as it should be, similarly to the liquid-gas transition. In this case, the contribution from $N^-$ is not negligible as it is populated in the vicinity of the chiral phase transition. In contrast, $\chi_2^+$ becomes negative and diverges at the critical point of the chiral phase transition. This is a direct consequence of the mass formula, which admits a minimum for $N_+$ at a finite value of $\sigma$ and the coefficient of the divergent $\pp \sigma/ \pp \mu_B$ becomes negative. We note that, albeit our results are obtained under the mean-field approximation, the inclusion of quantum fluctuations within the functional renormalization group (FRG) approach qualitatively preserves the same in-medium behavior of the baryon masses~\cite{Tripolt:2021jtp}.

\begin{figure}
    \centering
    \includegraphics[width=\linewidth]{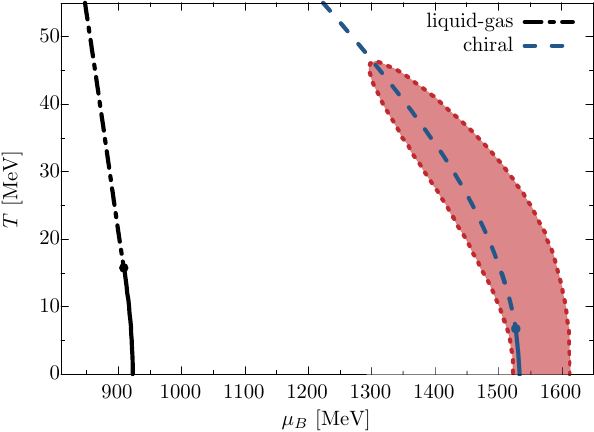}
    \caption{Low-temperature part of the phase diagram. Shown are the liquid-gas (black solid/dashed-dotted line) and chiral (blue solid/dashed line) phase transition/crossover lines. Circles indicate critical points below which the transitions are of the first order. The red envelope shows the region where $\chi_2^+$ is negative.}
    \label{fig:phase_diagram}
\end{figure}

In Fig.~\ref{fig:phase_diagram}, we show the low-temperature part of the phase diagram as a function of baryon chemical potential. At zero temperature, the system undergoes first-order liquid-gas and chiral phase transitions. With increasing temperature, both phase transitions develop their critical points above which there are no sharp transitions and they continue as smooth crossovers. At high temperature, they come closer together and finally merge~\cite{Sasaki:2010bp}. The red, dotted envelope marks the region, where the susceptibility $\chi_2^+$ becomes negative. This happens at $T=45~$MeV. Therefore, negative $\chi_2^+$ fluctuations signal approaching the region near the critical point of the chiral phase transition. We note that outside of this region, $\chi_2^+$ stays positive at all values of temperature and baryon chemical potential. Clearly, our study illustrates that the effects of chiral-criticality can become manifest differently in the properties of the fluctuations, depending on which degrees of freedom would be thermodynamically activated near the second-order phase transition at all temperatures and chemical potentials. 
We emphasize that one should also expect to see qualitative differences between higher-order fluctuations of positive- and negative-parity states. This is because they are proportional to higher-order derivatives $ \chi_n^\pm \sim \partial^n m_\pm / \partial \sigma^n$. Therefore, it is essential to utilize a framework with a self-consistent treatment of the chiral in-medium effects for a reliable description of the fluctuations of conserved charges.

To summarize, we have established for the first time the contribution to net-baryon fluctuations from baryonic chiral partners of opposite parity. We find that previously unexplored inclusion of the nucleon's chiral partner in a systematic way leads to a qualitative change in the structure of the fluctuations in the vicinity of the critical point. Namely, the critical fluctuations due to positive- and negative-parity baryonic chiral partners still carry the same critical exponents, however, with different coefficients and signs, such that their sum shows the expected critical scaling behavior of net-baryon number due to long-range correlations. Our conclusion is based on an assumption of mean-field dynamics, where we can split contributions from positive- and negative-parity states. Nevertheless, this allows indicating that the assumption about net-proton fluctuations being a good proxy for net-baryon fluctuations is not necessarily correct and requires further study.

\section{Conclusions}\label{sec:summary}

We have studied the qualitative structure of the fluctuations of conserved charges at finite density, focusing on the chiral-critical properties of the nucleon parity doublet. Utilizing the parity doublet model in the mean-field approximation, we have analyzed the second-order generalized susceptibilities of the net-baryon number density in the vicinity of the nuclear liquid-gas and chiral phase transitions. 

Our results are based on an assumption of mean-field dynamics, where the generalized susceptibilities are expressed as a sum of contributions from different species. This allowed us to consistently delineate the contributions from positive- and negative-parity chiral partners to the fluctuations of the net-baryon density. As expected, we have found that the fluctuations of the positive-parity state dominate the contribution in the vicinity of the liquid-gas phase transition, and its second-order susceptibility increases as the critical point is approached from high temperature. Contrary, the second-order susceptibility of the positive-parity state turns negative in the vicinity of the first-order chiral phase transition and diverges negatively at its critical point. At the same time, the susceptibility of the negative-parity state stays positive at all values of temperature and baryon chemical potential. This qualitative difference is traced back to the mass modification of parity doublet due to in-medium chiral effects. One possible next step beyond mean-field approximation would be the inclusion of mesonic fluctuations within the functional renormalization group (FRG) approach. Interestingly, the FRG results qualitatively preserve the same in-medium behavior of the baryon masses~\cite{Tripolt:2021jtp}. If the fluctuations of negative-parity states would turn out to be dominant over various beyond-mean-field correlations, one should expect to see deviations from the net-proton to net-baryon correspondence.

The qualitative differences in the sign of the positive- and negative-parity state fluctuations can also be useful in searching for possible critical points in the QCD phase diagram. In particular, our results bring significant and nontrivial differences of the critical behavior of the net-proton fluctuations in the vicinity of the liquid-gas and chiral phase transitions. This strongly suggests that in order to fully interpret the critical properties of the matter created in heavy-ion collisions, especially in the forthcoming large-scale nuclear experiments FAIR at GSI and NICA in Dubna, it is essential to consistently incorporate and understand the chiral in-medium effects carried by the parity doublers.

Our results revealed that in dense baryonic matter, the effects of chiral-criticality can manifest themselves differently in the properties of the fluctuations depending on the parity of baryons. It is conceivable that remnants of such effects can be present also at finite temperature and small or vanishing chemical potential. Consequently, measuring properties of net-proton number fluctuations in high-energy heavy-ion collisions is not necessarily sufficient to fully describe criticality due to chiral phase transition or its remnant. Further study to identify the role of chiral symmetry restoration and other hadronic interactions on the properties of fluctuations of parity doublets is in progress and will be reported elsewhere. 

\section*{Acknowledgements}
This work is supported partly by the Polish National Science Centre (NCN) under OPUS Grant No. 2018/31/B/ST2/01663 (K.R. and C.S.), Preludium Grant No. 2017/27/N/ST2/01973 (M.M.), and the program Excellence Initiative–Research University of the University of Wroc\l{}aw of the Ministry of Education and Science (M.M.). The work of C.S. was supported in part by the World Premier International Research Center Initiative (WPI) through MEXT, Japan. K.R. also acknowledges the support of the Polish
Ministry of Science and Higher Education.

\bibliographystyle{apsrev4-1}
\bibliography{biblio}

\begin{thebibliography}{85}%
\makeatletter
\providecommand \@ifxundefined [1]{%
 \@ifx{#1\undefined}
}%
\providecommand \@ifnum [1]{%
 \ifnum #1\expandafter \@firstoftwo
 \else \expandafter \@secondoftwo
 \fi
}%
\providecommand \@ifx [1]{%
 \ifx #1\expandafter \@firstoftwo
 \else \expandafter \@secondoftwo
 \fi
}%
\providecommand \natexlab [1]{#1}%
\providecommand \enquote  [1]{``#1''}%
\providecommand \bibnamefont  [1]{#1}%
\providecommand \bibfnamefont [1]{#1}%
\providecommand \citenamefont [1]{#1}%
\providecommand \href@noop [0]{\@secondoftwo}%
\providecommand \href [0]{\begingroup \@sanitize@url \@href}%
\providecommand \@href[1]{\@@startlink{#1}\@@href}%
\providecommand \@@href[1]{\endgroup#1\@@endlink}%
\providecommand \@sanitize@url [0]{\catcode `\\12\catcode `\$12\catcode
  `\&12\catcode `\#12\catcode `\^12\catcode `\_12\catcode `\%12\relax}%
\providecommand \@@startlink[1]{}%
\providecommand \@@endlink[0]{}%
\providecommand \url  [0]{\begingroup\@sanitize@url \@url }%
\providecommand \@url [1]{\endgroup\@href {#1}{\urlprefix }}%
\providecommand \urlprefix  [0]{URL }%
\providecommand \Eprint [0]{\href }%
\providecommand \doibase [0]{http://dx.doi.org/}%
\providecommand \selectlanguage [0]{\@gobble}%
\providecommand \bibinfo  [0]{\@secondoftwo}%
\providecommand \bibfield  [0]{\@secondoftwo}%
\providecommand \translation [1]{[#1]}%
\providecommand \BibitemOpen [0]{}%
\providecommand \bibitemStop [0]{}%
\providecommand \bibitemNoStop [0]{.\EOS\space}%
\providecommand \EOS [0]{\spacefactor3000\relax}%
\providecommand \BibitemShut  [1]{\csname bibitem#1\endcsname}%
\let\auto@bib@innerbib\@empty
\bibitem [{\citenamefont {Bazavov}\ \emph {et~al.}(2014)\citenamefont {Bazavov}
  \emph {et~al.}}]{Bazavov:2014pvz}%
  \BibitemOpen
  \bibfield  {author} {\bibinfo {author} {\bibfnamefont {A.}~\bibnamefont
  {Bazavov}} \emph {et~al.} (\bibinfo {collaboration} {HotQCD}),\ }\href
  {\doibase 10.1103/PhysRevD.90.094503} {\bibfield  {journal} {\bibinfo
  {journal} {Phys. Rev.}\ }\textbf {\bibinfo {volume} {D90}},\ \bibinfo {pages}
  {094503} (\bibinfo {year} {2014})},\ \Eprint {http://arxiv.org/abs/1407.6387}
  {arXiv:1407.6387 [hep-lat]} \BibitemShut {NoStop}%
\bibitem [{\citenamefont {Borsanyi}\ \emph {et~al.}(2018)\citenamefont
  {Borsanyi}, \citenamefont {Fodor}, \citenamefont {Guenther}, \citenamefont
  {Katz}, \citenamefont {Szabo}, \citenamefont {Pasztor}, \citenamefont
  {Portillo},\ and\ \citenamefont {Ratti}}]{Borsanyi:2018grb}%
  \BibitemOpen
  \bibfield  {author} {\bibinfo {author} {\bibfnamefont {S.}~\bibnamefont
  {Borsanyi}}, \bibinfo {author} {\bibfnamefont {Z.}~\bibnamefont {Fodor}},
  \bibinfo {author} {\bibfnamefont {J.~N.}\ \bibnamefont {Guenther}}, \bibinfo
  {author} {\bibfnamefont {S.~K.}\ \bibnamefont {Katz}}, \bibinfo {author}
  {\bibfnamefont {K.~K.}\ \bibnamefont {Szabo}}, \bibinfo {author}
  {\bibfnamefont {A.}~\bibnamefont {Pasztor}}, \bibinfo {author} {\bibfnamefont
  {I.}~\bibnamefont {Portillo}}, \ and\ \bibinfo {author} {\bibfnamefont
  {C.}~\bibnamefont {Ratti}},\ }\href {\doibase 10.1007/JHEP10(2018)205}
  {\bibfield  {journal} {\bibinfo  {journal} {JHEP}\ }\textbf {\bibinfo
  {volume} {10}},\ \bibinfo {pages} {205} (\bibinfo {year} {2018})},\ \Eprint
  {http://arxiv.org/abs/1805.04445} {arXiv:1805.04445 [hep-lat]} \BibitemShut
  {NoStop}%
\bibitem [{\citenamefont {Bazavov}\ \emph {et~al.}(2017)\citenamefont {Bazavov}
  \emph {et~al.}}]{Bazavov:2017dus}%
  \BibitemOpen
  \bibfield  {author} {\bibinfo {author} {\bibfnamefont {A.}~\bibnamefont
  {Bazavov}} \emph {et~al.},\ }\href {\doibase 10.1103/PhysRevD.95.054504}
  {\bibfield  {journal} {\bibinfo  {journal} {Phys. Rev. D}\ }\textbf {\bibinfo
  {volume} {95}},\ \bibinfo {pages} {054504} (\bibinfo {year} {2017})},\
  \Eprint {http://arxiv.org/abs/1701.04325} {arXiv:1701.04325 [hep-lat]}
  \BibitemShut {NoStop}%
\bibitem [{\citenamefont {Bazavov}\ \emph {et~al.}(2020)\citenamefont {Bazavov}
  \emph {et~al.}}]{Bazavov:2020bjn}%
  \BibitemOpen
  \bibfield  {author} {\bibinfo {author} {\bibfnamefont {A.}~\bibnamefont
  {Bazavov}} \emph {et~al.},\ }\href {\doibase 10.1103/PhysRevD.101.074502}
  {\bibfield  {journal} {\bibinfo  {journal} {Phys. Rev. D}\ }\textbf {\bibinfo
  {volume} {101}},\ \bibinfo {pages} {074502} (\bibinfo {year} {2020})},\
  \Eprint {http://arxiv.org/abs/2001.08530} {arXiv:2001.08530 [hep-lat]}
  \BibitemShut {NoStop}%
\bibitem [{\citenamefont {Aoki}\ \emph {et~al.}(2006)\citenamefont {Aoki},
  \citenamefont {Endrodi}, \citenamefont {Fodor}, \citenamefont {Katz},\ and\
  \citenamefont {Szabo}}]{Aoki:2006we}%
  \BibitemOpen
  \bibfield  {author} {\bibinfo {author} {\bibfnamefont {Y.}~\bibnamefont
  {Aoki}}, \bibinfo {author} {\bibfnamefont {G.}~\bibnamefont {Endrodi}},
  \bibinfo {author} {\bibfnamefont {Z.}~\bibnamefont {Fodor}}, \bibinfo
  {author} {\bibfnamefont {S.}~\bibnamefont {Katz}}, \ and\ \bibinfo {author}
  {\bibfnamefont {K.}~\bibnamefont {Szabo}},\ }\href {\doibase
  10.1038/nature05120} {\bibfield  {journal} {\bibinfo  {journal} {Nature}\
  }\textbf {\bibinfo {volume} {443}},\ \bibinfo {pages} {675} (\bibinfo {year}
  {2006})},\ \Eprint {http://arxiv.org/abs/hep-lat/0611014}
  {arXiv:hep-lat/0611014} \BibitemShut {NoStop}%
\bibitem [{\citenamefont {Bazavov}\ \emph {et~al.}(2019)\citenamefont {Bazavov}
  \emph {et~al.}}]{Bazavov:2018mes}%
  \BibitemOpen
  \bibfield  {author} {\bibinfo {author} {\bibfnamefont {A.}~\bibnamefont
  {Bazavov}} \emph {et~al.} (\bibinfo {collaboration} {HotQCD}),\ }\href
  {\doibase 10.1016/j.physletb.2019.05.013} {\bibfield  {journal} {\bibinfo
  {journal} {Phys.\ Lett.\ B}\ }\textbf {\bibinfo {volume} {795}},\ \bibinfo
  {pages} {15} (\bibinfo {year} {2019})},\ \Eprint
  {http://arxiv.org/abs/1812.08235} {arXiv:1812.08235 [hep-lat]} \BibitemShut
  {NoStop}%
\bibitem [{\citenamefont {Stephanov}\ \emph {et~al.}(1999)\citenamefont
  {Stephanov}, \citenamefont {Rajagopal},\ and\ \citenamefont
  {Shuryak}}]{Stephanov:1999zu}%
  \BibitemOpen
  \bibfield  {author} {\bibinfo {author} {\bibfnamefont {M.~A.}\ \bibnamefont
  {Stephanov}}, \bibinfo {author} {\bibfnamefont {K.}~\bibnamefont
  {Rajagopal}}, \ and\ \bibinfo {author} {\bibfnamefont {E.~V.}\ \bibnamefont
  {Shuryak}},\ }\href {\doibase 10.1103/PhysRevD.60.114028} {\bibfield
  {journal} {\bibinfo  {journal} {Phys. Rev. D}\ }\textbf {\bibinfo {volume}
  {60}},\ \bibinfo {pages} {114028} (\bibinfo {year} {1999})},\ \Eprint
  {http://arxiv.org/abs/hep-ph/9903292} {arXiv:hep-ph/9903292} \BibitemShut
  {NoStop}%
\bibitem [{\citenamefont {Asakawa}\ \emph {et~al.}(2000)\citenamefont
  {Asakawa}, \citenamefont {Heinz},\ and\ \citenamefont
  {Muller}}]{Asakawa:2000wh}%
  \BibitemOpen
  \bibfield  {author} {\bibinfo {author} {\bibfnamefont {M.}~\bibnamefont
  {Asakawa}}, \bibinfo {author} {\bibfnamefont {U.~W.}\ \bibnamefont {Heinz}},
  \ and\ \bibinfo {author} {\bibfnamefont {B.}~\bibnamefont {Muller}},\ }\href
  {\doibase 10.1103/PhysRevLett.85.2072} {\bibfield  {journal} {\bibinfo
  {journal} {Phys. Rev. Lett.}\ }\textbf {\bibinfo {volume} {85}},\ \bibinfo
  {pages} {2072} (\bibinfo {year} {2000})},\ \Eprint
  {http://arxiv.org/abs/hep-ph/0003169} {arXiv:hep-ph/0003169} \BibitemShut
  {NoStop}%
\bibitem [{\citenamefont {Hatta}\ and\ \citenamefont
  {Stephanov}(2003)}]{Hatta:2003wn}%
  \BibitemOpen
  \bibfield  {author} {\bibinfo {author} {\bibfnamefont {Y.}~\bibnamefont
  {Hatta}}\ and\ \bibinfo {author} {\bibfnamefont {M.}~\bibnamefont
  {Stephanov}},\ }\href {\doibase 10.1103/PhysRevLett.91.102003} {\bibfield
  {journal} {\bibinfo  {journal} {Phys. Rev. Lett.}\ }\textbf {\bibinfo
  {volume} {91}},\ \bibinfo {pages} {102003} (\bibinfo {year} {2003})},\
  \bibinfo {note} {[Erratum: Phys.Rev.Lett. 91, 129901 (2003)]},\ \Eprint
  {http://arxiv.org/abs/hep-ph/0302002} {arXiv:hep-ph/0302002} \BibitemShut
  {NoStop}%
\bibitem [{\citenamefont {Friman}\ \emph {et~al.}(2011)\citenamefont {Friman},
  \citenamefont {Karsch}, \citenamefont {Redlich},\ and\ \citenamefont
  {Skokov}}]{Friman:2011pf}%
  \BibitemOpen
  \bibfield  {author} {\bibinfo {author} {\bibfnamefont {B.}~\bibnamefont
  {Friman}}, \bibinfo {author} {\bibfnamefont {F.}~\bibnamefont {Karsch}},
  \bibinfo {author} {\bibfnamefont {K.}~\bibnamefont {Redlich}}, \ and\
  \bibinfo {author} {\bibfnamefont {V.}~\bibnamefont {Skokov}},\ }\href
  {\doibase 10.1140/epjc/s10052-011-1694-2} {\bibfield  {journal} {\bibinfo
  {journal} {Eur. Phys. J. C}\ }\textbf {\bibinfo {volume} {71}},\ \bibinfo
  {pages} {1694} (\bibinfo {year} {2011})},\ \Eprint
  {http://arxiv.org/abs/1103.3511} {arXiv:1103.3511 [hep-ph]} \BibitemShut
  {NoStop}%
\bibitem [{\citenamefont {Bazavov}\ \emph {et~al.}(2012)\citenamefont {Bazavov}
  \emph {et~al.}}]{Bazavov:2012vg}%
  \BibitemOpen
  \bibfield  {author} {\bibinfo {author} {\bibfnamefont {A.}~\bibnamefont
  {Bazavov}} \emph {et~al.},\ }\href {\doibase 10.1103/PhysRevLett.109.192302}
  {\bibfield  {journal} {\bibinfo  {journal} {Phys. Rev. Lett.}\ }\textbf
  {\bibinfo {volume} {109}},\ \bibinfo {pages} {192302} (\bibinfo {year}
  {2012})},\ \Eprint {http://arxiv.org/abs/1208.1220} {arXiv:1208.1220
  [hep-lat]} \BibitemShut {NoStop}%
\bibitem [{\citenamefont {Borsanyi}\ \emph {et~al.}(2014)\citenamefont
  {Borsanyi}, \citenamefont {Fodor}, \citenamefont {Katz}, \citenamefont
  {Krieg}, \citenamefont {Ratti},\ and\ \citenamefont
  {Szabo}}]{Borsanyi:2014ewa}%
  \BibitemOpen
  \bibfield  {author} {\bibinfo {author} {\bibfnamefont {S.}~\bibnamefont
  {Borsanyi}}, \bibinfo {author} {\bibfnamefont {Z.}~\bibnamefont {Fodor}},
  \bibinfo {author} {\bibfnamefont {S.}~\bibnamefont {Katz}}, \bibinfo {author}
  {\bibfnamefont {S.}~\bibnamefont {Krieg}}, \bibinfo {author} {\bibfnamefont
  {C.}~\bibnamefont {Ratti}}, \ and\ \bibinfo {author} {\bibfnamefont
  {K.}~\bibnamefont {Szabo}},\ }\href {\doibase 10.1103/PhysRevLett.113.052301}
  {\bibfield  {journal} {\bibinfo  {journal} {Phys. Rev. Lett.}\ }\textbf
  {\bibinfo {volume} {113}},\ \bibinfo {pages} {052301} (\bibinfo {year}
  {2014})},\ \Eprint {http://arxiv.org/abs/1403.4576} {arXiv:1403.4576
  [hep-lat]} \BibitemShut {NoStop}%
\bibitem [{\citenamefont {Karsch}\ and\ \citenamefont
  {Redlich}(2011)}]{Karsch:2010ck}%
  \BibitemOpen
  \bibfield  {author} {\bibinfo {author} {\bibfnamefont {F.}~\bibnamefont
  {Karsch}}\ and\ \bibinfo {author} {\bibfnamefont {K.}~\bibnamefont
  {Redlich}},\ }\href {\doibase 10.1016/j.physletb.2010.10.046} {\bibfield
  {journal} {\bibinfo  {journal} {Phys.\ Lett.\ B}\ }\textbf {\bibinfo {volume}
  {695}},\ \bibinfo {pages} {136} (\bibinfo {year} {2011})},\ \Eprint
  {http://arxiv.org/abs/1007.2581} {arXiv:1007.2581 [hep-ph]} \BibitemShut
  {NoStop}%
\bibitem [{\citenamefont {Braun-Munzinger}\ \emph {et~al.}(2015)\citenamefont
  {Braun-Munzinger}, \citenamefont {Kalweit}, \citenamefont {Redlich},\ and\
  \citenamefont {Stachel}}]{Braun-Munzinger:2014lba}%
  \BibitemOpen
  \bibfield  {author} {\bibinfo {author} {\bibfnamefont {P.}~\bibnamefont
  {Braun-Munzinger}}, \bibinfo {author} {\bibfnamefont {A.}~\bibnamefont
  {Kalweit}}, \bibinfo {author} {\bibfnamefont {K.}~\bibnamefont {Redlich}}, \
  and\ \bibinfo {author} {\bibfnamefont {J.}~\bibnamefont {Stachel}},\ }\href
  {\doibase 10.1016/j.physletb.2015.05.077} {\bibfield  {journal} {\bibinfo
  {journal} {Phys. Lett. B}\ }\textbf {\bibinfo {volume} {747}},\ \bibinfo
  {pages} {292} (\bibinfo {year} {2015})},\ \Eprint
  {http://arxiv.org/abs/1412.8614} {arXiv:1412.8614 [hep-ph]} \BibitemShut
  {NoStop}%
\bibitem [{\citenamefont {Vovchenko}\ \emph {et~al.}(2020)\citenamefont
  {Vovchenko}, \citenamefont {Savchuk}, \citenamefont {Poberezhnyuk},
  \citenamefont {Gorenstein},\ and\ \citenamefont {Koch}}]{Vovchenko:2020tsr}%
  \BibitemOpen
  \bibfield  {author} {\bibinfo {author} {\bibfnamefont {V.}~\bibnamefont
  {Vovchenko}}, \bibinfo {author} {\bibfnamefont {O.}~\bibnamefont {Savchuk}},
  \bibinfo {author} {\bibfnamefont {R.~V.}\ \bibnamefont {Poberezhnyuk}},
  \bibinfo {author} {\bibfnamefont {M.~I.}\ \bibnamefont {Gorenstein}}, \ and\
  \bibinfo {author} {\bibfnamefont {V.}~\bibnamefont {Koch}},\ }\href {\doibase
  10.1016/j.physletb.2020.135868} {\bibfield  {journal} {\bibinfo  {journal}
  {Phys. Lett. B}\ }\textbf {\bibinfo {volume} {811}},\ \bibinfo {pages}
  {135868} (\bibinfo {year} {2020})},\ \Eprint
  {http://arxiv.org/abs/2003.13905} {arXiv:2003.13905 [hep-ph]} \BibitemShut
  {NoStop}%
\bibitem [{\citenamefont {Braun-Munzinger}\ \emph {et~al.}(2021)\citenamefont
  {Braun-Munzinger}, \citenamefont {Friman}, \citenamefont {Redlich},
  \citenamefont {Rustamov},\ and\ \citenamefont
  {Stachel}}]{Braun-Munzinger:2020jbk}%
  \BibitemOpen
  \bibfield  {author} {\bibinfo {author} {\bibfnamefont {P.}~\bibnamefont
  {Braun-Munzinger}}, \bibinfo {author} {\bibfnamefont {B.}~\bibnamefont
  {Friman}}, \bibinfo {author} {\bibfnamefont {K.}~\bibnamefont {Redlich}},
  \bibinfo {author} {\bibfnamefont {A.}~\bibnamefont {Rustamov}}, \ and\
  \bibinfo {author} {\bibfnamefont {J.}~\bibnamefont {Stachel}},\ }\href
  {\doibase 10.1016/j.nuclphysa.2021.122141} {\bibfield  {journal} {\bibinfo
  {journal} {Nucl. Phys. A}\ }\textbf {\bibinfo {volume} {1008}},\ \bibinfo
  {pages} {122141} (\bibinfo {year} {2021})},\ \Eprint
  {http://arxiv.org/abs/2007.02463} {arXiv:2007.02463 [nucl-th]} \BibitemShut
  {NoStop}%
\bibitem [{\citenamefont {Stephanov}(2011)}]{Stephanov:2011pb}%
  \BibitemOpen
  \bibfield  {author} {\bibinfo {author} {\bibfnamefont {M.}~\bibnamefont
  {Stephanov}},\ }\href {\doibase 10.1103/PhysRevLett.107.052301} {\bibfield
  {journal} {\bibinfo  {journal} {Phys. Rev. Lett.}\ }\textbf {\bibinfo
  {volume} {107}},\ \bibinfo {pages} {052301} (\bibinfo {year} {2011})},\
  \Eprint {http://arxiv.org/abs/1104.1627} {arXiv:1104.1627 [hep-ph]}
  \BibitemShut {NoStop}%
\bibitem [{\citenamefont {Karsch}(2019)}]{Karsch:2019mbv}%
  \BibitemOpen
  \bibfield  {author} {\bibinfo {author} {\bibfnamefont {F.}~\bibnamefont
  {Karsch}},\ }\href {\doibase 10.22323/1.347.0163} {\bibfield  {journal}
  {\bibinfo  {journal} {PoS}\ }\textbf {\bibinfo {volume} {CORFU2018}},\
  \bibinfo {pages} {163} (\bibinfo {year} {2019})},\ \Eprint
  {http://arxiv.org/abs/1905.03936} {arXiv:1905.03936 [hep-lat]} \BibitemShut
  {NoStop}%
\bibitem [{\citenamefont {Braun-Munzinger}\ \emph {et~al.}(2017)\citenamefont
  {Braun-Munzinger}, \citenamefont {Rustamov},\ and\ \citenamefont
  {Stachel}}]{Braun-Munzinger:2016yjz}%
  \BibitemOpen
  \bibfield  {author} {\bibinfo {author} {\bibfnamefont {P.}~\bibnamefont
  {Braun-Munzinger}}, \bibinfo {author} {\bibfnamefont {A.}~\bibnamefont
  {Rustamov}}, \ and\ \bibinfo {author} {\bibfnamefont {J.}~\bibnamefont
  {Stachel}},\ }\href {\doibase 10.1016/j.nuclphysa.2017.01.011} {\bibfield
  {journal} {\bibinfo  {journal} {Nucl. Phys. A}\ }\textbf {\bibinfo {volume}
  {960}},\ \bibinfo {pages} {114} (\bibinfo {year} {2017})},\ \Eprint
  {http://arxiv.org/abs/1612.00702} {arXiv:1612.00702 [nucl-th]} \BibitemShut
  {NoStop}%
\bibitem [{\citenamefont {Aggarwal}\ \emph {et~al.}(2010)\citenamefont
  {Aggarwal} \emph {et~al.}}]{STAR2010}%
  \BibitemOpen
  \bibfield  {author} {\bibinfo {author} {\bibfnamefont {M.~M.}\ \bibnamefont
  {Aggarwal}} \emph {et~al.} (\bibinfo {collaboration} {STAR}),\ }\href@noop {}
  {\  (\bibinfo {year} {2010})},\ \Eprint {http://arxiv.org/abs/1007.2613}
  {arXiv:1007.2613 [nucl-ex]} \BibitemShut {NoStop}%
\bibitem [{\citenamefont
  {Ma\'ckowiak-Paw\l{}owska}(2021)}]{Mackowiak-Pawlowska:2020glz}%
  \BibitemOpen
  \bibfield  {author} {\bibinfo {author} {\bibfnamefont {M.}~\bibnamefont
  {Ma\'ckowiak-Paw\l{}owska}} (\bibinfo {collaboration} {NA61/SHINE}),\ }\href
  {\doibase 10.1016/j.nuclphysa.2020.121753} {\bibfield  {journal} {\bibinfo
  {journal} {Nucl. Phys. A}\ }\textbf {\bibinfo {volume} {1005}},\ \bibinfo
  {pages} {121753} (\bibinfo {year} {2021})},\ \Eprint
  {http://arxiv.org/abs/2002.04847} {arXiv:2002.04847 [nucl-ex]} \BibitemShut
  {NoStop}%
\bibitem [{\citenamefont {{Braun-Munzinger}}\ \emph {et~al.}(2004)\citenamefont
  {{Braun-Munzinger}}, \citenamefont {{Redlich}},\ and\ \citenamefont
  {{Stachel}}}]{BraunMunzinger:2003zd}%
  \BibitemOpen
  \bibfield  {author} {\bibinfo {author} {\bibfnamefont {P.}~\bibnamefont
  {{Braun-Munzinger}}}, \bibinfo {author} {\bibfnamefont {K.}~\bibnamefont
  {{Redlich}}}, \ and\ \bibinfo {author} {\bibfnamefont {J.}~\bibnamefont
  {{Stachel}}},\ }\href {\doibase 10.1142/9789812795533_0008} {\emph {\bibinfo
  {title} {Quark-Gluon Plasma 3. Edited by Hwa Rudolph \& Wang Xin-Nian.
  Published by World Scientific Publishing Co. Pte. Ltd}}}\ (\bibinfo {year}
  {2004})\ pp.\ \bibinfo {pages} {491--599},\ \Eprint
  {http://arxiv.org/abs/nucl-th/0304013} {arXiv:nucl-th/0304013} \BibitemShut
  {NoStop}%
\bibitem [{\citenamefont {Andronic}\ \emph {et~al.}(2018)\citenamefont
  {Andronic}, \citenamefont {Braun-Munzinger}, \citenamefont {Redlich},\ and\
  \citenamefont {Stachel}}]{Andronic:2017pug}%
  \BibitemOpen
  \bibfield  {author} {\bibinfo {author} {\bibfnamefont {A.}~\bibnamefont
  {Andronic}}, \bibinfo {author} {\bibfnamefont {P.}~\bibnamefont
  {Braun-Munzinger}}, \bibinfo {author} {\bibfnamefont {K.}~\bibnamefont
  {Redlich}}, \ and\ \bibinfo {author} {\bibfnamefont {J.}~\bibnamefont
  {Stachel}},\ }\href {\doibase 10.1038/s41586-018-0491-6} {\bibfield
  {journal} {\bibinfo  {journal} {Nature}\ }\textbf {\bibinfo {volume} {561}},\
  \bibinfo {pages} {321} (\bibinfo {year} {2018})},\ \Eprint
  {http://arxiv.org/abs/1710.09425} {arXiv:1710.09425 [nucl-th]} \BibitemShut
  {NoStop}%
\bibitem [{\citenamefont {Venugopalan}\ and\ \citenamefont
  {Prakash}(1992)}]{Venugopalan:1992hy}%
  \BibitemOpen
  \bibfield  {author} {\bibinfo {author} {\bibfnamefont {R.}~\bibnamefont
  {Venugopalan}}\ and\ \bibinfo {author} {\bibfnamefont {M.}~\bibnamefont
  {Prakash}},\ }\href {\doibase 10.1016/0375-9474(92)90005-5} {\bibfield
  {journal} {\bibinfo  {journal} {Nucl. Phys. A}\ }\textbf {\bibinfo {volume}
  {546}},\ \bibinfo {pages} {718} (\bibinfo {year} {1992})}\BibitemShut
  {NoStop}%
\bibitem [{\citenamefont {Broniowski}\ \emph {et~al.}(2015)\citenamefont
  {Broniowski}, \citenamefont {Giacosa},\ and\ \citenamefont
  {Begun}}]{Broniowski:2015oha}%
  \BibitemOpen
  \bibfield  {author} {\bibinfo {author} {\bibfnamefont {W.}~\bibnamefont
  {Broniowski}}, \bibinfo {author} {\bibfnamefont {F.}~\bibnamefont {Giacosa}},
  \ and\ \bibinfo {author} {\bibfnamefont {V.}~\bibnamefont {Begun}},\ }\href
  {\doibase 10.1103/PhysRevC.92.034905} {\bibfield  {journal} {\bibinfo
  {journal} {Phys. Rev. C}\ }\textbf {\bibinfo {volume} {92}},\ \bibinfo
  {pages} {034905} (\bibinfo {year} {2015})},\ \Eprint
  {http://arxiv.org/abs/1506.01260} {arXiv:1506.01260 [nucl-th]} \BibitemShut
  {NoStop}%
\bibitem [{\citenamefont {Friman}\ \emph {et~al.}(2015)\citenamefont {Friman},
  \citenamefont {Lo}, \citenamefont {Marczenko}, \citenamefont {Redlich},\ and\
  \citenamefont {Sasaki}}]{Friman:2015zua}%
  \BibitemOpen
  \bibfield  {author} {\bibinfo {author} {\bibfnamefont {B.}~\bibnamefont
  {Friman}}, \bibinfo {author} {\bibfnamefont {P.~M.}\ \bibnamefont {Lo}},
  \bibinfo {author} {\bibfnamefont {M.}~\bibnamefont {Marczenko}}, \bibinfo
  {author} {\bibfnamefont {K.}~\bibnamefont {Redlich}}, \ and\ \bibinfo
  {author} {\bibfnamefont {C.}~\bibnamefont {Sasaki}},\ }\href {\doibase
  10.1103/PhysRevD.92.074003} {\bibfield  {journal} {\bibinfo  {journal} {Phys.
  Rev. D}\ }\textbf {\bibinfo {volume} {92}},\ \bibinfo {pages} {074003}
  (\bibinfo {year} {2015})},\ \Eprint {http://arxiv.org/abs/1507.04183}
  {arXiv:1507.04183 [hep-ph]} \BibitemShut {NoStop}%
\bibitem [{\citenamefont {Huovinen}\ \emph {et~al.}(2017)\citenamefont
  {Huovinen}, \citenamefont {Lo}, \citenamefont {Marczenko}, \citenamefont
  {Morita}, \citenamefont {Redlich},\ and\ \citenamefont
  {Sasaki}}]{Huovinen:2016xxq}%
  \BibitemOpen
  \bibfield  {author} {\bibinfo {author} {\bibfnamefont {P.}~\bibnamefont
  {Huovinen}}, \bibinfo {author} {\bibfnamefont {P.~M.}\ \bibnamefont {Lo}},
  \bibinfo {author} {\bibfnamefont {M.}~\bibnamefont {Marczenko}}, \bibinfo
  {author} {\bibfnamefont {K.}~\bibnamefont {Morita}}, \bibinfo {author}
  {\bibfnamefont {K.}~\bibnamefont {Redlich}}, \ and\ \bibinfo {author}
  {\bibfnamefont {C.}~\bibnamefont {Sasaki}},\ }\href {\doibase
  10.1016/j.physletb.2017.03.060} {\bibfield  {journal} {\bibinfo  {journal}
  {Phys. Lett. B}\ }\textbf {\bibinfo {volume} {769}},\ \bibinfo {pages} {509}
  (\bibinfo {year} {2017})},\ \Eprint {http://arxiv.org/abs/1608.06817}
  {arXiv:1608.06817 [hep-ph]} \BibitemShut {NoStop}%
\bibitem [{\citenamefont {Lo}\ \emph {et~al.}(2018)\citenamefont {Lo},
  \citenamefont {Friman}, \citenamefont {Redlich},\ and\ \citenamefont
  {Sasaki}}]{Lo:2017lym}%
  \BibitemOpen
  \bibfield  {author} {\bibinfo {author} {\bibfnamefont {P.~M.}\ \bibnamefont
  {Lo}}, \bibinfo {author} {\bibfnamefont {B.}~\bibnamefont {Friman}}, \bibinfo
  {author} {\bibfnamefont {K.}~\bibnamefont {Redlich}}, \ and\ \bibinfo
  {author} {\bibfnamefont {C.}~\bibnamefont {Sasaki}},\ }\href {\doibase
  10.1016/j.physletb.2018.01.016} {\bibfield  {journal} {\bibinfo  {journal}
  {Phys. Lett. B}\ }\textbf {\bibinfo {volume} {778}},\ \bibinfo {pages} {454}
  (\bibinfo {year} {2018})},\ \Eprint {http://arxiv.org/abs/1710.02711}
  {arXiv:1710.02711 [hep-ph]} \BibitemShut {NoStop}%
\bibitem [{\citenamefont {Majumder}\ and\ \citenamefont
  {Muller}(2010)}]{Majumder:2010ik}%
  \BibitemOpen
  \bibfield  {author} {\bibinfo {author} {\bibfnamefont {A.}~\bibnamefont
  {Majumder}}\ and\ \bibinfo {author} {\bibfnamefont {B.}~\bibnamefont
  {Muller}},\ }\href {\doibase 10.1103/PhysRevLett.105.252002} {\bibfield
  {journal} {\bibinfo  {journal} {Phys. Rev. Lett.}\ }\textbf {\bibinfo
  {volume} {105}},\ \bibinfo {pages} {252002} (\bibinfo {year} {2010})},\
  \Eprint {http://arxiv.org/abs/1008.1747} {arXiv:1008.1747 [hep-ph]}
  \BibitemShut {NoStop}%
\bibitem [{\citenamefont {Andronic}\ \emph {et~al.}(2012)\citenamefont
  {Andronic}, \citenamefont {Braun-Munzinger}, \citenamefont {Stachel},\ and\
  \citenamefont {Winn}}]{Andronic:2012ut}%
  \BibitemOpen
  \bibfield  {author} {\bibinfo {author} {\bibfnamefont {A.}~\bibnamefont
  {Andronic}}, \bibinfo {author} {\bibfnamefont {P.}~\bibnamefont
  {Braun-Munzinger}}, \bibinfo {author} {\bibfnamefont {J.}~\bibnamefont
  {Stachel}}, \ and\ \bibinfo {author} {\bibfnamefont {M.}~\bibnamefont
  {Winn}},\ }\href {\doibase 10.1016/j.physletb.2012.10.001} {\bibfield
  {journal} {\bibinfo  {journal} {Phys. Lett. B}\ }\textbf {\bibinfo {volume}
  {718}},\ \bibinfo {pages} {80} (\bibinfo {year} {2012})},\ \Eprint
  {http://arxiv.org/abs/1201.0693} {arXiv:1201.0693 [nucl-th]} \BibitemShut
  {NoStop}%
\bibitem [{\citenamefont {Albright}\ \emph {et~al.}(2014)\citenamefont
  {Albright}, \citenamefont {Kapusta},\ and\ \citenamefont
  {Young}}]{Albright:2014gva}%
  \BibitemOpen
  \bibfield  {author} {\bibinfo {author} {\bibfnamefont {M.}~\bibnamefont
  {Albright}}, \bibinfo {author} {\bibfnamefont {J.}~\bibnamefont {Kapusta}}, \
  and\ \bibinfo {author} {\bibfnamefont {C.}~\bibnamefont {Young}},\ }\href
  {\doibase 10.1103/PhysRevC.90.024915} {\bibfield  {journal} {\bibinfo
  {journal} {Phys. Rev. C}\ }\textbf {\bibinfo {volume} {90}},\ \bibinfo
  {pages} {024915} (\bibinfo {year} {2014})},\ \Eprint
  {http://arxiv.org/abs/1404.7540} {arXiv:1404.7540 [nucl-th]} \BibitemShut
  {NoStop}%
\bibitem [{\citenamefont {Vovchenko}\ \emph {et~al.}(2015)\citenamefont
  {Vovchenko}, \citenamefont {Anchishkin},\ and\ \citenamefont
  {Gorenstein}}]{Vovchenko:2014pka}%
  \BibitemOpen
  \bibfield  {author} {\bibinfo {author} {\bibfnamefont {V.}~\bibnamefont
  {Vovchenko}}, \bibinfo {author} {\bibfnamefont {D.}~\bibnamefont
  {Anchishkin}}, \ and\ \bibinfo {author} {\bibfnamefont {M.}~\bibnamefont
  {Gorenstein}},\ }\href {\doibase 10.1103/PhysRevC.91.024905} {\bibfield
  {journal} {\bibinfo  {journal} {Phys. Rev. C}\ }\textbf {\bibinfo {volume}
  {91}},\ \bibinfo {pages} {024905} (\bibinfo {year} {2015})},\ \Eprint
  {http://arxiv.org/abs/1412.5478} {arXiv:1412.5478 [nucl-th]} \BibitemShut
  {NoStop}%
\bibitem [{\citenamefont {Lo}\ \emph {et~al.}(2015)\citenamefont {Lo},
  \citenamefont {Marczenko}, \citenamefont {Redlich},\ and\ \citenamefont
  {Sasaki}}]{Lo:2015cca}%
  \BibitemOpen
  \bibfield  {author} {\bibinfo {author} {\bibfnamefont {P.~M.}\ \bibnamefont
  {Lo}}, \bibinfo {author} {\bibfnamefont {M.}~\bibnamefont {Marczenko}},
  \bibinfo {author} {\bibfnamefont {K.}~\bibnamefont {Redlich}}, \ and\
  \bibinfo {author} {\bibfnamefont {C.}~\bibnamefont {Sasaki}},\ }\href
  {\doibase 10.1103/PhysRevC.92.055206} {\bibfield  {journal} {\bibinfo
  {journal} {Phys. Rev. C}\ }\textbf {\bibinfo {volume} {92}},\ \bibinfo
  {pages} {055206} (\bibinfo {year} {2015})},\ \Eprint
  {http://arxiv.org/abs/1507.06398} {arXiv:1507.06398 [nucl-th]} \BibitemShut
  {NoStop}%
\bibitem [{\citenamefont {Man~Lo}\ \emph {et~al.}(2016)\citenamefont {Man~Lo},
  \citenamefont {Marczenko}, \citenamefont {Redlich},\ and\ \citenamefont
  {Sasaki}}]{ManLo:2016pgd}%
  \BibitemOpen
  \bibfield  {author} {\bibinfo {author} {\bibfnamefont {P.}~\bibnamefont
  {Man~Lo}}, \bibinfo {author} {\bibfnamefont {M.}~\bibnamefont {Marczenko}},
  \bibinfo {author} {\bibfnamefont {K.}~\bibnamefont {Redlich}}, \ and\
  \bibinfo {author} {\bibfnamefont {C.}~\bibnamefont {Sasaki}},\ }\href
  {\doibase 10.1140/epja/i2016-16235-6} {\bibfield  {journal} {\bibinfo
  {journal} {Eur. Phys. J. A}\ }\textbf {\bibinfo {volume} {52}},\ \bibinfo
  {pages} {235} (\bibinfo {year} {2016})}\BibitemShut {NoStop}%
\bibitem [{\citenamefont {Andronic}\ \emph {et~al.}(2021)\citenamefont
  {Andronic}, \citenamefont {Braun-Munzinger}, \citenamefont {G\"und\"uz},
  \citenamefont {Kirchhoff}, \citenamefont {K\"ohler}, \citenamefont
  {Stachel},\ and\ \citenamefont {Winn}}]{Andronic:2020iyg}%
  \BibitemOpen
  \bibfield  {author} {\bibinfo {author} {\bibfnamefont {A.}~\bibnamefont
  {Andronic}}, \bibinfo {author} {\bibfnamefont {P.}~\bibnamefont
  {Braun-Munzinger}}, \bibinfo {author} {\bibfnamefont {D.}~\bibnamefont
  {G\"und\"uz}}, \bibinfo {author} {\bibfnamefont {Y.}~\bibnamefont
  {Kirchhoff}}, \bibinfo {author} {\bibfnamefont {M.~K.}\ \bibnamefont
  {K\"ohler}}, \bibinfo {author} {\bibfnamefont {J.}~\bibnamefont {Stachel}}, \
  and\ \bibinfo {author} {\bibfnamefont {M.}~\bibnamefont {Winn}},\ }\href
  {\doibase 10.1016/j.nuclphysa.2021.122176} {\bibfield  {journal} {\bibinfo
  {journal} {Nucl. Phys. A}\ }\textbf {\bibinfo {volume} {1010}},\ \bibinfo
  {pages} {122176} (\bibinfo {year} {2021})},\ \Eprint
  {http://arxiv.org/abs/2011.03826} {arXiv:2011.03826 [nucl-th]} \BibitemShut
  {NoStop}%
\bibitem [{\citenamefont {Vovchenko}\ \emph {et~al.}(2017)\citenamefont
  {Vovchenko}, \citenamefont {Gorenstein},\ and\ \citenamefont
  {Stoecker}}]{Vovchenko:2016rkn}%
  \BibitemOpen
  \bibfield  {author} {\bibinfo {author} {\bibfnamefont {V.}~\bibnamefont
  {Vovchenko}}, \bibinfo {author} {\bibfnamefont {M.~I.}\ \bibnamefont
  {Gorenstein}}, \ and\ \bibinfo {author} {\bibfnamefont {H.}~\bibnamefont
  {Stoecker}},\ }\href {\doibase 10.1103/PhysRevLett.118.182301} {\bibfield
  {journal} {\bibinfo  {journal} {Phys. Rev. Lett.}\ }\textbf {\bibinfo
  {volume} {118}},\ \bibinfo {pages} {182301} (\bibinfo {year} {2017})},\
  \Eprint {http://arxiv.org/abs/1609.03975} {arXiv:1609.03975 [hep-ph]}
  \BibitemShut {NoStop}%
\bibitem [{\citenamefont {Marczenko}\ \emph {et~al.}(2021)\citenamefont
  {Marczenko}, \citenamefont {Redlich},\ and\ \citenamefont
  {Sasaki}}]{Marczenko:2021icv}%
  \BibitemOpen
  \bibfield  {author} {\bibinfo {author} {\bibfnamefont {M.}~\bibnamefont
  {Marczenko}}, \bibinfo {author} {\bibfnamefont {K.}~\bibnamefont {Redlich}},
  \ and\ \bibinfo {author} {\bibfnamefont {C.}~\bibnamefont {Sasaki}},\ }\href
  {\doibase 10.1103/physrevd.103.054035} {\bibfield  {journal} {\bibinfo
  {journal} {Phys. Rev. D}\ }\textbf {\bibinfo {volume} {103}},\ \bibinfo
  {pages} {054035} (\bibinfo {year} {2021})}\BibitemShut {NoStop}%
\bibitem [{\citenamefont {Aarts}\ \emph {et~al.}(2015)\citenamefont {Aarts},
  \citenamefont {Allton}, \citenamefont {Hands}, \citenamefont {Jäger},
  \citenamefont {Praki},\ and\ \citenamefont {Skullerud}}]{Aarts:2015mma}%
  \BibitemOpen
  \bibfield  {author} {\bibinfo {author} {\bibfnamefont {G.}~\bibnamefont
  {Aarts}}, \bibinfo {author} {\bibfnamefont {C.}~\bibnamefont {Allton}},
  \bibinfo {author} {\bibfnamefont {S.}~\bibnamefont {Hands}}, \bibinfo
  {author} {\bibfnamefont {B.}~\bibnamefont {Jäger}}, \bibinfo {author}
  {\bibfnamefont {C.}~\bibnamefont {Praki}}, \ and\ \bibinfo {author}
  {\bibfnamefont {J.-I.}\ \bibnamefont {Skullerud}},\ }\href {\doibase
  10.1103/PhysRevD.92.014503} {\bibfield  {journal} {\bibinfo  {journal} {Phys.
  Rev.}\ }\textbf {\bibinfo {volume} {D92}},\ \bibinfo {pages} {014503}
  (\bibinfo {year} {2015})},\ \Eprint {http://arxiv.org/abs/1502.03603}
  {arXiv:1502.03603 [hep-lat]} \BibitemShut {NoStop}%
\bibitem [{\citenamefont {Aarts}\ \emph {et~al.}(2017)\citenamefont {Aarts},
  \citenamefont {Allton}, \citenamefont {De~Boni}, \citenamefont {Hands},
  \citenamefont {J\"ager}, \citenamefont {Praki},\ and\ \citenamefont
  {Skullerud}}]{Aarts:2017rrl}%
  \BibitemOpen
  \bibfield  {author} {\bibinfo {author} {\bibfnamefont {G.}~\bibnamefont
  {Aarts}}, \bibinfo {author} {\bibfnamefont {C.}~\bibnamefont {Allton}},
  \bibinfo {author} {\bibfnamefont {D.}~\bibnamefont {De~Boni}}, \bibinfo
  {author} {\bibfnamefont {S.}~\bibnamefont {Hands}}, \bibinfo {author}
  {\bibfnamefont {B.}~\bibnamefont {J\"ager}}, \bibinfo {author} {\bibfnamefont
  {C.}~\bibnamefont {Praki}}, \ and\ \bibinfo {author} {\bibfnamefont {J.-I.}\
  \bibnamefont {Skullerud}},\ }\href {\doibase 10.1007/JHEP06(2017)034}
  {\bibfield  {journal} {\bibinfo  {journal} {JHEP}\ }\textbf {\bibinfo
  {volume} {06}},\ \bibinfo {pages} {034} (\bibinfo {year} {2017})},\ \Eprint
  {http://arxiv.org/abs/1703.09246} {arXiv:1703.09246 [hep-lat]} \BibitemShut
  {NoStop}%
\bibitem [{\citenamefont {Aarts}\ \emph {et~al.}(2019)\citenamefont {Aarts},
  \citenamefont {Allton}, \citenamefont {De~Boni},\ and\ \citenamefont
  {Jäger}}]{Aarts:2018glk}%
  \BibitemOpen
  \bibfield  {author} {\bibinfo {author} {\bibfnamefont {G.}~\bibnamefont
  {Aarts}}, \bibinfo {author} {\bibfnamefont {C.}~\bibnamefont {Allton}},
  \bibinfo {author} {\bibfnamefont {D.}~\bibnamefont {De~Boni}}, \ and\
  \bibinfo {author} {\bibfnamefont {B.}~\bibnamefont {Jäger}},\ }\href
  {\doibase 10.1103/PhysRevD.99.074503} {\bibfield  {journal} {\bibinfo
  {journal} {Phys. Rev.}\ }\textbf {\bibinfo {volume} {D99}},\ \bibinfo {pages}
  {074503} (\bibinfo {year} {2019})},\ \Eprint
  {http://arxiv.org/abs/1812.07393} {arXiv:1812.07393 [hep-lat]} \BibitemShut
  {NoStop}%
\bibitem [{\citenamefont {{De~Tar}}\ and\ \citenamefont
  {Kunihiro}(1989)}]{Detar:1988kn}%
  \BibitemOpen
  \bibfield  {author} {\bibinfo {author} {\bibfnamefont {C.~E.}\ \bibnamefont
  {{De~Tar}}}\ and\ \bibinfo {author} {\bibfnamefont {T.}~\bibnamefont
  {Kunihiro}},\ }\href {\doibase 10.1103/PhysRevD.39.2805} {\bibfield
  {journal} {\bibinfo  {journal} {Phys. Rev.}\ }\textbf {\bibinfo {volume}
  {D39}},\ \bibinfo {pages} {2805} (\bibinfo {year} {1989})}\BibitemShut
  {NoStop}%
\bibitem [{\citenamefont {Jido}\ \emph {et~al.}(2000)\citenamefont {Jido},
  \citenamefont {Hatsuda},\ and\ \citenamefont {Kunihiro}}]{Jido:1999hd}%
  \BibitemOpen
  \bibfield  {author} {\bibinfo {author} {\bibfnamefont {D.}~\bibnamefont
  {Jido}}, \bibinfo {author} {\bibfnamefont {T.}~\bibnamefont {Hatsuda}}, \
  and\ \bibinfo {author} {\bibfnamefont {T.}~\bibnamefont {Kunihiro}},\ }\href
  {\doibase 10.1103/PhysRevLett.84.3252} {\bibfield  {journal} {\bibinfo
  {journal} {Phys. Rev. Lett.}\ }\textbf {\bibinfo {volume} {84}},\ \bibinfo
  {pages} {3252} (\bibinfo {year} {2000})},\ \Eprint
  {http://arxiv.org/abs/hep-ph/9910375} {arXiv:hep-ph/9910375 [hep-ph]}
  \BibitemShut {NoStop}%
\bibitem [{\citenamefont {Jido}\ \emph {et~al.}(2001)\citenamefont {Jido},
  \citenamefont {Oka},\ and\ \citenamefont {Hosaka}}]{Jido:2001nt}%
  \BibitemOpen
  \bibfield  {author} {\bibinfo {author} {\bibfnamefont {D.}~\bibnamefont
  {Jido}}, \bibinfo {author} {\bibfnamefont {M.}~\bibnamefont {Oka}}, \ and\
  \bibinfo {author} {\bibfnamefont {A.}~\bibnamefont {Hosaka}},\ }\href
  {\doibase 10.1143/PTP.106.873} {\bibfield  {journal} {\bibinfo  {journal}
  {Prog. Theor. Phys.}\ }\textbf {\bibinfo {volume} {106}},\ \bibinfo {pages}
  {873} (\bibinfo {year} {2001})},\ \Eprint
  {http://arxiv.org/abs/hep-ph/0110005} {arXiv:hep-ph/0110005 [hep-ph]}
  \BibitemShut {NoStop}%
\bibitem [{\citenamefont {Dexheimer}\ \emph {et~al.}(2008)\citenamefont
  {Dexheimer}, \citenamefont {Schramm},\ and\ \citenamefont
  {Zschiesche}}]{Dexheimer:2007tn}%
  \BibitemOpen
  \bibfield  {author} {\bibinfo {author} {\bibfnamefont {V.}~\bibnamefont
  {Dexheimer}}, \bibinfo {author} {\bibfnamefont {S.}~\bibnamefont {Schramm}},
  \ and\ \bibinfo {author} {\bibfnamefont {D.}~\bibnamefont {Zschiesche}},\
  }\href {\doibase 10.1103/PhysRevC.77.025803} {\bibfield  {journal} {\bibinfo
  {journal} {Phys. Rev.}\ }\textbf {\bibinfo {volume} {C77}},\ \bibinfo {pages}
  {025803} (\bibinfo {year} {2008})},\ \Eprint {http://arxiv.org/abs/0710.4192}
  {arXiv:0710.4192 [nucl-th]} \BibitemShut {NoStop}%
\bibitem [{\citenamefont {Gallas}\ \emph {et~al.}(2010)\citenamefont {Gallas},
  \citenamefont {Giacosa},\ and\ \citenamefont {Rischke}}]{Gallas:2009qp}%
  \BibitemOpen
  \bibfield  {author} {\bibinfo {author} {\bibfnamefont {S.}~\bibnamefont
  {Gallas}}, \bibinfo {author} {\bibfnamefont {F.}~\bibnamefont {Giacosa}}, \
  and\ \bibinfo {author} {\bibfnamefont {D.~H.}\ \bibnamefont {Rischke}},\
  }\href {\doibase 10.1103/PhysRevD.82.014004} {\bibfield  {journal} {\bibinfo
  {journal} {Phys. Rev.}\ }\textbf {\bibinfo {volume} {D82}},\ \bibinfo {pages}
  {014004} (\bibinfo {year} {2010})},\ \Eprint {http://arxiv.org/abs/0907.5084}
  {arXiv:0907.5084 [hep-ph]} \BibitemShut {NoStop}%
\bibitem [{\citenamefont {Paeng}\ \emph {et~al.}(2012)\citenamefont {Paeng},
  \citenamefont {Lee}, \citenamefont {Rho},\ and\ \citenamefont
  {Sasaki}}]{Paeng:2011hy}%
  \BibitemOpen
  \bibfield  {author} {\bibinfo {author} {\bibfnamefont {W.-G.}\ \bibnamefont
  {Paeng}}, \bibinfo {author} {\bibfnamefont {H.~K.}\ \bibnamefont {Lee}},
  \bibinfo {author} {\bibfnamefont {M.}~\bibnamefont {Rho}}, \ and\ \bibinfo
  {author} {\bibfnamefont {C.}~\bibnamefont {Sasaki}},\ }\href {\doibase
  10.1103/PhysRevD.85.054022} {\bibfield  {journal} {\bibinfo  {journal} {Phys.
  Rev.}\ }\textbf {\bibinfo {volume} {D85}},\ \bibinfo {pages} {054022}
  (\bibinfo {year} {2012})},\ \Eprint {http://arxiv.org/abs/1109.5431}
  {arXiv:1109.5431 [hep-ph]} \BibitemShut {NoStop}%
\bibitem [{\citenamefont {Sasaki}\ \emph {et~al.}(2011)\citenamefont {Sasaki},
  \citenamefont {Lee}, \citenamefont {Paeng},\ and\ \citenamefont
  {Rho}}]{Sasaki:2011ff}%
  \BibitemOpen
  \bibfield  {author} {\bibinfo {author} {\bibfnamefont {C.}~\bibnamefont
  {Sasaki}}, \bibinfo {author} {\bibfnamefont {H.~K.}\ \bibnamefont {Lee}},
  \bibinfo {author} {\bibfnamefont {W.-G.}\ \bibnamefont {Paeng}}, \ and\
  \bibinfo {author} {\bibfnamefont {M.}~\bibnamefont {Rho}},\ }\href {\doibase
  10.1103/PhysRevD.84.034011} {\bibfield  {journal} {\bibinfo  {journal} {Phys.
  Rev.}\ }\textbf {\bibinfo {volume} {D84}},\ \bibinfo {pages} {034011}
  (\bibinfo {year} {2011})},\ \Eprint {http://arxiv.org/abs/1103.0184}
  {arXiv:1103.0184 [hep-ph]} \BibitemShut {NoStop}%
\bibitem [{\citenamefont {Gallas}\ \emph {et~al.}(2011)\citenamefont {Gallas},
  \citenamefont {Giacosa},\ and\ \citenamefont {Pagliara}}]{Gallas:2011qp}%
  \BibitemOpen
  \bibfield  {author} {\bibinfo {author} {\bibfnamefont {S.}~\bibnamefont
  {Gallas}}, \bibinfo {author} {\bibfnamefont {F.}~\bibnamefont {Giacosa}}, \
  and\ \bibinfo {author} {\bibfnamefont {G.}~\bibnamefont {Pagliara}},\ }\href
  {\doibase 10.1016/j.nuclphysa.2011.09.008} {\bibfield  {journal} {\bibinfo
  {journal} {Nucl. Phys. A}\ }\textbf {\bibinfo {volume} {872}},\ \bibinfo
  {pages} {13} (\bibinfo {year} {2011})},\ \Eprint
  {http://arxiv.org/abs/1105.5003} {arXiv:1105.5003 [hep-ph]} \BibitemShut
  {NoStop}%
\bibitem [{\citenamefont {Zschiesche}\ \emph {et~al.}(2007)\citenamefont
  {Zschiesche}, \citenamefont {Tolos}, \citenamefont {Schaffner-Bielich},\ and\
  \citenamefont {Pisarski}}]{Zschiesche:2006zj}%
  \BibitemOpen
  \bibfield  {author} {\bibinfo {author} {\bibfnamefont {D.}~\bibnamefont
  {Zschiesche}}, \bibinfo {author} {\bibfnamefont {L.}~\bibnamefont {Tolos}},
  \bibinfo {author} {\bibfnamefont {J.}~\bibnamefont {Schaffner-Bielich}}, \
  and\ \bibinfo {author} {\bibfnamefont {R.~D.}\ \bibnamefont {Pisarski}},\
  }\href {\doibase 10.1103/PhysRevC.75.055202} {\bibfield  {journal} {\bibinfo
  {journal} {Phys. Rev.}\ }\textbf {\bibinfo {volume} {C75}},\ \bibinfo {pages}
  {055202} (\bibinfo {year} {2007})},\ \Eprint
  {http://arxiv.org/abs/nucl-th/0608044} {arXiv:nucl-th/0608044 [nucl-th]}
  \BibitemShut {NoStop}%
\bibitem [{\citenamefont {Benic}\ \emph {et~al.}(2015)\citenamefont {Benic},
  \citenamefont {Mishustin},\ and\ \citenamefont {Sasaki}}]{Benic:2015pia}%
  \BibitemOpen
  \bibfield  {author} {\bibinfo {author} {\bibfnamefont {S.}~\bibnamefont
  {Benic}}, \bibinfo {author} {\bibfnamefont {I.}~\bibnamefont {Mishustin}}, \
  and\ \bibinfo {author} {\bibfnamefont {C.}~\bibnamefont {Sasaki}},\ }\href
  {\doibase 10.1103/PhysRevD.91.125034} {\bibfield  {journal} {\bibinfo
  {journal} {Phys. Rev.}\ }\textbf {\bibinfo {volume} {D91}},\ \bibinfo {pages}
  {125034} (\bibinfo {year} {2015})},\ \Eprint
  {http://arxiv.org/abs/1502.05969} {arXiv:1502.05969 [hep-ph]} \BibitemShut
  {NoStop}%
\bibitem [{\citenamefont {Marczenko}\ and\ \citenamefont
  {Sasaki}(2018)}]{Marczenko:2017huu}%
  \BibitemOpen
  \bibfield  {author} {\bibinfo {author} {\bibfnamefont {M.}~\bibnamefont
  {Marczenko}}\ and\ \bibinfo {author} {\bibfnamefont {C.}~\bibnamefont
  {Sasaki}},\ }\href {\doibase 10.1103/PhysRevD.97.036011} {\bibfield
  {journal} {\bibinfo  {journal} {Phys. Rev.}\ }\textbf {\bibinfo {volume}
  {D97}},\ \bibinfo {pages} {036011} (\bibinfo {year} {2018})},\ \Eprint
  {http://arxiv.org/abs/1711.05521} {arXiv:1711.05521 [hep-ph]} \BibitemShut
  {NoStop}%
\bibitem [{\citenamefont {Marczenko}\ \emph {et~al.}(2018)\citenamefont
  {Marczenko}, \citenamefont {Blaschke}, \citenamefont {Redlich},\ and\
  \citenamefont {Sasaki}}]{Marczenko:2018jui}%
  \BibitemOpen
  \bibfield  {author} {\bibinfo {author} {\bibfnamefont {M.}~\bibnamefont
  {Marczenko}}, \bibinfo {author} {\bibfnamefont {D.}~\bibnamefont {Blaschke}},
  \bibinfo {author} {\bibfnamefont {K.}~\bibnamefont {Redlich}}, \ and\
  \bibinfo {author} {\bibfnamefont {C.}~\bibnamefont {Sasaki}},\ }\href
  {\doibase 10.1103/PhysRevD.98.103021} {\bibfield  {journal} {\bibinfo
  {journal} {Phys. Rev.}\ }\textbf {\bibinfo {volume} {D98}},\ \bibinfo {pages}
  {103021} (\bibinfo {year} {2018})},\ \Eprint
  {http://arxiv.org/abs/1805.06886} {arXiv:1805.06886} \BibitemShut {NoStop}%
\bibitem [{\citenamefont {Marczenko}\ \emph {et~al.}(2019)\citenamefont
  {Marczenko}, \citenamefont {Blaschke}, \citenamefont {Redlich},\ and\
  \citenamefont {Sasaki}}]{Marczenko:2019trv}%
  \BibitemOpen
  \bibfield  {author} {\bibinfo {author} {\bibfnamefont {M.}~\bibnamefont
  {Marczenko}}, \bibinfo {author} {\bibfnamefont {D.}~\bibnamefont {Blaschke}},
  \bibinfo {author} {\bibfnamefont {K.}~\bibnamefont {Redlich}}, \ and\
  \bibinfo {author} {\bibfnamefont {C.}~\bibnamefont {Sasaki}},\ }\href
  {\doibase 10.3390/universe5080180} {\bibfield  {journal} {\bibinfo  {journal}
  {Universe}\ }\textbf {\bibinfo {volume} {5}},\ \bibinfo {pages} {180}
  (\bibinfo {year} {2019})},\ \Eprint {http://arxiv.org/abs/1905.04974}
  {arXiv:1905.04974 [nucl-th]} \BibitemShut {NoStop}%
\bibitem [{\citenamefont {Marczenko}(2020)}]{Marczenko:2020wlc}%
  \BibitemOpen
  \bibfield  {author} {\bibinfo {author} {\bibfnamefont {M.}~\bibnamefont
  {Marczenko}},\ }\href {\doibase 10.1140/epjst/e2020-000093-3} {\bibfield
  {journal} {\bibinfo  {journal} {Eur. Phys. J. ST}\ }\textbf {\bibinfo
  {volume} {229}},\ \bibinfo {pages} {3651} (\bibinfo {year} {2020})},\ \Eprint
  {http://arxiv.org/abs/2005.14535} {arXiv:2005.14535 [nucl-th]} \BibitemShut
  {NoStop}%
\bibitem [{\citenamefont {Marczenko}\ \emph {et~al.}(2020)\citenamefont
  {Marczenko}, \citenamefont {Blaschke}, \citenamefont {Redlich},\ and\
  \citenamefont {Sasaki}}]{Marczenko:2020jma}%
  \BibitemOpen
  \bibfield  {author} {\bibinfo {author} {\bibfnamefont {M.}~\bibnamefont
  {Marczenko}}, \bibinfo {author} {\bibfnamefont {D.}~\bibnamefont {Blaschke}},
  \bibinfo {author} {\bibfnamefont {K.}~\bibnamefont {Redlich}}, \ and\
  \bibinfo {author} {\bibfnamefont {C.}~\bibnamefont {Sasaki}},\ }\href
  {\doibase 10.1051/0004-6361/202038211} {\bibfield  {journal} {\bibinfo
  {journal} {Astron. Astrophys.}\ }\textbf {\bibinfo {volume} {643}},\ \bibinfo
  {pages} {A82} (\bibinfo {year} {2020})},\ \Eprint
  {http://arxiv.org/abs/2004.09566} {arXiv:2004.09566 [astro-ph.HE]}
  \BibitemShut {NoStop}%
\bibitem [{\citenamefont {Marczenko}\ \emph
  {et~al.}(2022{\natexlab{a}})\citenamefont {Marczenko}, \citenamefont
  {Redlich},\ and\ \citenamefont {Sasaki}}]{Marczenko:2021uaj}%
  \BibitemOpen
  \bibfield  {author} {\bibinfo {author} {\bibfnamefont {M.}~\bibnamefont
  {Marczenko}}, \bibinfo {author} {\bibfnamefont {K.}~\bibnamefont {Redlich}},
  \ and\ \bibinfo {author} {\bibfnamefont {C.}~\bibnamefont {Sasaki}},\ }\href
  {\doibase 10.3847/2041-8213/ac4b61} {\bibfield  {journal} {\bibinfo
  {journal} {Astrophys. J. Lett.}\ }\textbf {\bibinfo {volume} {925}},\
  \bibinfo {pages} {L23} (\bibinfo {year} {2022}{\natexlab{a}})},\ \Eprint
  {http://arxiv.org/abs/2110.11056} {arXiv:2110.11056 [nucl-th]} \BibitemShut
  {NoStop}%
\bibitem [{\citenamefont {Marczenko}\ \emph
  {et~al.}(2022{\natexlab{b}})\citenamefont {Marczenko}, \citenamefont
  {Redlich},\ and\ \citenamefont {Sasaki}}]{Marczenko:2022hyt}%
  \BibitemOpen
  \bibfield  {author} {\bibinfo {author} {\bibfnamefont {M.}~\bibnamefont
  {Marczenko}}, \bibinfo {author} {\bibfnamefont {K.}~\bibnamefont {Redlich}},
  \ and\ \bibinfo {author} {\bibfnamefont {C.}~\bibnamefont {Sasaki}},\ }\href
  {\doibase 10.1103/PhysRevD.105.103009} {\bibfield  {journal} {\bibinfo
  {journal} {Phys. Rev. D}\ }\textbf {\bibinfo {volume} {105}},\ \bibinfo
  {pages} {103009} (\bibinfo {year} {2022}{\natexlab{b}})},\ \Eprint
  {http://arxiv.org/abs/2203.00269} {arXiv:2203.00269 [nucl-th]} \BibitemShut
  {NoStop}%
\bibitem [{\citenamefont {Mukherjee}\ \emph
  {et~al.}(2017{\natexlab{a}})\citenamefont {Mukherjee}, \citenamefont
  {Schramm}, \citenamefont {Steinheimer},\ and\ \citenamefont
  {Dexheimer}}]{Mukherjee:2017jzi}%
  \BibitemOpen
  \bibfield  {author} {\bibinfo {author} {\bibfnamefont {A.}~\bibnamefont
  {Mukherjee}}, \bibinfo {author} {\bibfnamefont {S.}~\bibnamefont {Schramm}},
  \bibinfo {author} {\bibfnamefont {J.}~\bibnamefont {Steinheimer}}, \ and\
  \bibinfo {author} {\bibfnamefont {V.}~\bibnamefont {Dexheimer}},\ }\href
  {\doibase 10.1051/0004-6361/201731505} {\bibfield  {journal} {\bibinfo
  {journal} {Astron. Astrophys.}\ }\textbf {\bibinfo {volume} {608}},\ \bibinfo
  {pages} {A110} (\bibinfo {year} {2017}{\natexlab{a}})},\ \Eprint
  {http://arxiv.org/abs/1706.09191} {arXiv:1706.09191 [nucl-th]} \BibitemShut
  {NoStop}%
\bibitem [{\citenamefont {Mukherjee}\ \emph
  {et~al.}(2017{\natexlab{b}})\citenamefont {Mukherjee}, \citenamefont
  {Steinheimer},\ and\ \citenamefont {Schramm}}]{Mukherjee:2016nhb}%
  \BibitemOpen
  \bibfield  {author} {\bibinfo {author} {\bibfnamefont {A.}~\bibnamefont
  {Mukherjee}}, \bibinfo {author} {\bibfnamefont {J.}~\bibnamefont
  {Steinheimer}}, \ and\ \bibinfo {author} {\bibfnamefont {S.}~\bibnamefont
  {Schramm}},\ }\href {\doibase 10.1103/PhysRevC.96.025205} {\bibfield
  {journal} {\bibinfo  {journal} {Phys. Rev.}\ }\textbf {\bibinfo {volume}
  {C96}},\ \bibinfo {pages} {025205} (\bibinfo {year} {2017}{\natexlab{b}})},\
  \Eprint {http://arxiv.org/abs/1611.10144} {arXiv:1611.10144 [nucl-th]}
  \BibitemShut {NoStop}%
\bibitem [{\citenamefont {Dexheimer}\ \emph {et~al.}(2013)\citenamefont
  {Dexheimer}, \citenamefont {Steinheimer}, \citenamefont {Negreiros},\ and\
  \citenamefont {Schramm}}]{Dexheimer:2012eu}%
  \BibitemOpen
  \bibfield  {author} {\bibinfo {author} {\bibfnamefont {V.}~\bibnamefont
  {Dexheimer}}, \bibinfo {author} {\bibfnamefont {J.}~\bibnamefont
  {Steinheimer}}, \bibinfo {author} {\bibfnamefont {R.}~\bibnamefont
  {Negreiros}}, \ and\ \bibinfo {author} {\bibfnamefont {S.}~\bibnamefont
  {Schramm}},\ }\href {\doibase 10.1103/PhysRevC.87.015804} {\bibfield
  {journal} {\bibinfo  {journal} {Phys. Rev.}\ }\textbf {\bibinfo {volume}
  {C87}},\ \bibinfo {pages} {015804} (\bibinfo {year} {2013})},\ \Eprint
  {http://arxiv.org/abs/1206.3086} {arXiv:1206.3086 [astro-ph.HE]} \BibitemShut
  {NoStop}%
\bibitem [{\citenamefont {Steinheimer}\ \emph
  {et~al.}(2011{\natexlab{a}})\citenamefont {Steinheimer}, \citenamefont
  {Schramm},\ and\ \citenamefont {Stocker}}]{Steinheimer:2011ea}%
  \BibitemOpen
  \bibfield  {author} {\bibinfo {author} {\bibfnamefont {J.}~\bibnamefont
  {Steinheimer}}, \bibinfo {author} {\bibfnamefont {S.}~\bibnamefont
  {Schramm}}, \ and\ \bibinfo {author} {\bibfnamefont {H.}~\bibnamefont
  {Stocker}},\ }\href {\doibase 10.1103/PhysRevC.84.045208} {\bibfield
  {journal} {\bibinfo  {journal} {Phys. Rev.}\ }\textbf {\bibinfo {volume}
  {C84}},\ \bibinfo {pages} {045208} (\bibinfo {year} {2011}{\natexlab{a}})},\
  \Eprint {http://arxiv.org/abs/1108.2596} {arXiv:1108.2596 [hep-ph]}
  \BibitemShut {NoStop}%
\bibitem [{\citenamefont {Weyrich}\ \emph {et~al.}(2015)\citenamefont
  {Weyrich}, \citenamefont {Strodthoff},\ and\ \citenamefont {von
  Smekal}}]{Weyrich:2015hha}%
  \BibitemOpen
  \bibfield  {author} {\bibinfo {author} {\bibfnamefont {J.}~\bibnamefont
  {Weyrich}}, \bibinfo {author} {\bibfnamefont {N.}~\bibnamefont {Strodthoff}},
  \ and\ \bibinfo {author} {\bibfnamefont {L.}~\bibnamefont {von Smekal}},\
  }\href {\doibase 10.1103/PhysRevC.92.015214} {\bibfield  {journal} {\bibinfo
  {journal} {Phys. Rev.}\ }\textbf {\bibinfo {volume} {C92}},\ \bibinfo {pages}
  {015214} (\bibinfo {year} {2015})},\ \Eprint
  {http://arxiv.org/abs/1504.02697} {arXiv:1504.02697 [nucl-th]} \BibitemShut
  {NoStop}%
\bibitem [{\citenamefont {Sasaki}\ and\ \citenamefont
  {Mishustin}(2010)}]{Sasaki:2010bp}%
  \BibitemOpen
  \bibfield  {author} {\bibinfo {author} {\bibfnamefont {C.}~\bibnamefont
  {Sasaki}}\ and\ \bibinfo {author} {\bibfnamefont {I.}~\bibnamefont
  {Mishustin}},\ }\href {\doibase 10.1103/PhysRevC.82.035204} {\bibfield
  {journal} {\bibinfo  {journal} {Phys. Rev.}\ }\textbf {\bibinfo {volume}
  {C82}},\ \bibinfo {pages} {035204} (\bibinfo {year} {2010})},\ \Eprint
  {http://arxiv.org/abs/1005.4811} {arXiv:1005.4811 [hep-ph]} \BibitemShut
  {NoStop}%
\bibitem [{\citenamefont {Yamazaki}\ and\ \citenamefont
  {Harada}(2019{\natexlab{a}})}]{Yamazaki:2018stk}%
  \BibitemOpen
  \bibfield  {author} {\bibinfo {author} {\bibfnamefont {T.}~\bibnamefont
  {Yamazaki}}\ and\ \bibinfo {author} {\bibfnamefont {M.}~\bibnamefont
  {Harada}},\ }\href {\doibase 10.1103/PhysRevD.99.034012} {\bibfield
  {journal} {\bibinfo  {journal} {Phys. Rev. D}\ }\textbf {\bibinfo {volume}
  {99}},\ \bibinfo {pages} {034012} (\bibinfo {year} {2019}{\natexlab{a}})},\
  \Eprint {http://arxiv.org/abs/1809.02359} {arXiv:1809.02359 [hep-ph]}
  \BibitemShut {NoStop}%
\bibitem [{\citenamefont {Yamazaki}\ and\ \citenamefont
  {Harada}(2019{\natexlab{b}})}]{Yamazaki:2019tuo}%
  \BibitemOpen
  \bibfield  {author} {\bibinfo {author} {\bibfnamefont {T.}~\bibnamefont
  {Yamazaki}}\ and\ \bibinfo {author} {\bibfnamefont {M.}~\bibnamefont
  {Harada}},\ }\href {\doibase 10.1103/PhysRevC.100.025205} {\bibfield
  {journal} {\bibinfo  {journal} {Phys. Rev.}\ }\textbf {\bibinfo {volume}
  {C100}},\ \bibinfo {pages} {025205} (\bibinfo {year} {2019}{\natexlab{b}})},\
  \Eprint {http://arxiv.org/abs/1901.02167} {arXiv:1901.02167 [nucl-th]}
  \BibitemShut {NoStop}%
\bibitem [{\citenamefont {Ishikawa}\ \emph {et~al.}(2019)\citenamefont
  {Ishikawa}, \citenamefont {Nakayama},\ and\ \citenamefont
  {Suzuki}}]{Ishikawa:2018yey}%
  \BibitemOpen
  \bibfield  {author} {\bibinfo {author} {\bibfnamefont {T.}~\bibnamefont
  {Ishikawa}}, \bibinfo {author} {\bibfnamefont {K.}~\bibnamefont {Nakayama}},
  \ and\ \bibinfo {author} {\bibfnamefont {K.}~\bibnamefont {Suzuki}},\ }\href
  {\doibase 10.1103/PhysRevD.99.054010} {\bibfield  {journal} {\bibinfo
  {journal} {Phys. Rev.}\ }\textbf {\bibinfo {volume} {D99}},\ \bibinfo {pages}
  {054010} (\bibinfo {year} {2019})},\ \Eprint
  {http://arxiv.org/abs/1812.10964} {arXiv:1812.10964 [hep-ph]} \BibitemShut
  {NoStop}%
\bibitem [{\citenamefont {Steinheimer}\ \emph
  {et~al.}(2011{\natexlab{b}})\citenamefont {Steinheimer}, \citenamefont
  {Schramm},\ and\ \citenamefont {Stocker}}]{Steinheimer:2010ib}%
  \BibitemOpen
  \bibfield  {author} {\bibinfo {author} {\bibfnamefont {J.}~\bibnamefont
  {Steinheimer}}, \bibinfo {author} {\bibfnamefont {S.}~\bibnamefont
  {Schramm}}, \ and\ \bibinfo {author} {\bibfnamefont {H.}~\bibnamefont
  {Stocker}},\ }\href {\doibase 10.1088/0954-3899/38/3/035001} {\bibfield
  {journal} {\bibinfo  {journal} {J. Phys.}\ }\textbf {\bibinfo {volume}
  {G38}},\ \bibinfo {pages} {035001} (\bibinfo {year} {2011}{\natexlab{b}})},\
  \Eprint {http://arxiv.org/abs/1009.5239} {arXiv:1009.5239 [hep-ph]}
  \BibitemShut {NoStop}%
\bibitem [{\citenamefont {Giacosa}(2012)}]{Giacosa:2011qd}%
  \BibitemOpen
  \bibfield  {author} {\bibinfo {author} {\bibfnamefont {F.}~\bibnamefont
  {Giacosa}},\ }\href {\doibase 10.1016/j.ppnp.2011.12.039} {\bibfield
  {journal} {\bibinfo  {journal} {Prog. Part. Nucl. Phys.}\ }\textbf {\bibinfo
  {volume} {67}},\ \bibinfo {pages} {332} (\bibinfo {year} {2012})},\ \Eprint
  {http://arxiv.org/abs/1111.4944} {arXiv:1111.4944 [hep-ph]} \BibitemShut
  {NoStop}%
\bibitem [{\citenamefont {Motohiro}\ \emph {et~al.}(2015)\citenamefont
  {Motohiro}, \citenamefont {Kim},\ and\ \citenamefont
  {Harada}}]{Motohiro:2015taa}%
  \BibitemOpen
  \bibfield  {author} {\bibinfo {author} {\bibfnamefont {Y.}~\bibnamefont
  {Motohiro}}, \bibinfo {author} {\bibfnamefont {Y.}~\bibnamefont {Kim}}, \
  and\ \bibinfo {author} {\bibfnamefont {M.}~\bibnamefont {Harada}},\ }\href
  {\doibase 10.1103/PhysRevC.92.025201, 10.1103/PhysRevC.95.059903} {\bibfield
  {journal} {\bibinfo  {journal} {Phys. Rev.}\ }\textbf {\bibinfo {volume}
  {C92}},\ \bibinfo {pages} {025201} (\bibinfo {year} {2015})},\ \bibinfo
  {note} {[Erratum: Phys. Rev.C95,no.5,059903(2017)]},\ \Eprint
  {http://arxiv.org/abs/1505.00988} {arXiv:1505.00988 [nucl-th]} \BibitemShut
  {NoStop}%
\bibitem [{\citenamefont {Minamikawa}\ \emph {et~al.}(2021)\citenamefont
  {Minamikawa}, \citenamefont {Kojo},\ and\ \citenamefont
  {Harada}}]{Minamikawa:2020jfj}%
  \BibitemOpen
  \bibfield  {author} {\bibinfo {author} {\bibfnamefont {T.}~\bibnamefont
  {Minamikawa}}, \bibinfo {author} {\bibfnamefont {T.}~\bibnamefont {Kojo}}, \
  and\ \bibinfo {author} {\bibfnamefont {M.}~\bibnamefont {Harada}},\ }\href
  {\doibase 10.1103/PhysRevC.103.045205} {\bibfield  {journal} {\bibinfo
  {journal} {Phys. Rev. C}\ }\textbf {\bibinfo {volume} {103}},\ \bibinfo
  {pages} {045205} (\bibinfo {year} {2021})},\ \Eprint
  {http://arxiv.org/abs/2011.13684} {arXiv:2011.13684 [nucl-th]} \BibitemShut
  {NoStop}%
\bibitem [{\citenamefont {Gell-Mann}\ and\ \citenamefont
  {Levy}(1960)}]{GellMann:1960np}%
  \BibitemOpen
  \bibfield  {author} {\bibinfo {author} {\bibfnamefont {M.}~\bibnamefont
  {Gell-Mann}}\ and\ \bibinfo {author} {\bibfnamefont {M.}~\bibnamefont
  {Levy}},\ }\href {\doibase 10.1007/BF02859738} {\bibfield  {journal}
  {\bibinfo  {journal} {Nuovo Cim.}\ }\textbf {\bibinfo {volume} {16}},\
  \bibinfo {pages} {705} (\bibinfo {year} {1960})}\BibitemShut {NoStop}%
\bibitem [{\citenamefont {Serot}\ and\ \citenamefont
  {Walecka}(1986)}]{Serot:1984ey}%
  \BibitemOpen
  \bibfield  {author} {\bibinfo {author} {\bibfnamefont {B.~D.}\ \bibnamefont
  {Serot}}\ and\ \bibinfo {author} {\bibfnamefont {J.~D.}\ \bibnamefont
  {Walecka}},\ }\href@noop {} {\bibfield  {journal} {\bibinfo  {journal} {Adv.
  Nucl. Phys.}\ }\textbf {\bibinfo {volume} {16}},\ \bibinfo {pages} {1}
  (\bibinfo {year} {1986})}\BibitemShut {NoStop}%
\bibitem [{\citenamefont {Motornenko}\ \emph {et~al.}(2020)\citenamefont
  {Motornenko}, \citenamefont {Steinheimer}, \citenamefont {Vovchenko},
  \citenamefont {Schramm},\ and\ \citenamefont
  {Stoecker}}]{Motornenko:2019arp}%
  \BibitemOpen
  \bibfield  {author} {\bibinfo {author} {\bibfnamefont {A.}~\bibnamefont
  {Motornenko}}, \bibinfo {author} {\bibfnamefont {J.}~\bibnamefont
  {Steinheimer}}, \bibinfo {author} {\bibfnamefont {V.}~\bibnamefont
  {Vovchenko}}, \bibinfo {author} {\bibfnamefont {S.}~\bibnamefont {Schramm}},
  \ and\ \bibinfo {author} {\bibfnamefont {H.}~\bibnamefont {Stoecker}},\
  }\href {\doibase 10.1103/PhysRevC.101.034904} {\bibfield  {journal} {\bibinfo
   {journal} {Phys. Rev.}\ }\textbf {\bibinfo {volume} {C101}},\ \bibinfo
  {pages} {034904} (\bibinfo {year} {2020})},\ \Eprint
  {http://arxiv.org/abs/1905.00866} {arXiv:1905.00866 [hep-ph]} \BibitemShut
  {NoStop}%
\bibitem [{\citenamefont {Wetterich}(1993)}]{Wetterich:1992yh}%
  \BibitemOpen
  \bibfield  {author} {\bibinfo {author} {\bibfnamefont {C.}~\bibnamefont
  {Wetterich}},\ }\href {\doibase 10.1016/0370-2693(93)90726-X} {\bibfield
  {journal} {\bibinfo  {journal} {Phys. Lett. B}\ }\textbf {\bibinfo {volume}
  {301}},\ \bibinfo {pages} {90} (\bibinfo {year} {1993})},\ \Eprint
  {http://arxiv.org/abs/1710.05815} {arXiv:1710.05815 [hep-th]} \BibitemShut
  {NoStop}%
\bibitem [{\citenamefont {Morris}(1994)}]{Morris:1993qb}%
  \BibitemOpen
  \bibfield  {author} {\bibinfo {author} {\bibfnamefont {T.~R.}\ \bibnamefont
  {Morris}},\ }\href {\doibase 10.1142/S0217751X94000972} {\bibfield  {journal}
  {\bibinfo  {journal} {Int. J. Mod. Phys. A}\ }\textbf {\bibinfo {volume}
  {9}},\ \bibinfo {pages} {2411} (\bibinfo {year} {1994})},\ \Eprint
  {http://arxiv.org/abs/hep-ph/9308265} {arXiv:hep-ph/9308265} \BibitemShut
  {NoStop}%
\bibitem [{\citenamefont {Ellwanger}(1994)}]{Ellwanger:1993mw}%
  \BibitemOpen
  \bibfield  {author} {\bibinfo {author} {\bibfnamefont {U.}~\bibnamefont
  {Ellwanger}},\ }\href {\doibase 10.1007/BF01555911} {\bibfield  {journal}
  {\bibinfo  {journal} {Z. Phys. C}\ }\textbf {\bibinfo {volume} {62}},\
  \bibinfo {pages} {503} (\bibinfo {year} {1994})},\ \Eprint
  {http://arxiv.org/abs/hep-ph/9308260} {arXiv:hep-ph/9308260} \BibitemShut
  {NoStop}%
\bibitem [{\citenamefont {Berges}\ \emph {et~al.}(2002)\citenamefont {Berges},
  \citenamefont {Tetradis},\ and\ \citenamefont {Wetterich}}]{Berges:2000ew}%
  \BibitemOpen
  \bibfield  {author} {\bibinfo {author} {\bibfnamefont {J.}~\bibnamefont
  {Berges}}, \bibinfo {author} {\bibfnamefont {N.}~\bibnamefont {Tetradis}}, \
  and\ \bibinfo {author} {\bibfnamefont {C.}~\bibnamefont {Wetterich}},\ }\href
  {\doibase 10.1016/S0370-1573(01)00098-9} {\bibfield  {journal} {\bibinfo
  {journal} {Phys. Rept.}\ }\textbf {\bibinfo {volume} {363}},\ \bibinfo
  {pages} {223} (\bibinfo {year} {2002})},\ \Eprint
  {http://arxiv.org/abs/hep-ph/0005122} {arXiv:hep-ph/0005122} \BibitemShut
  {NoStop}%
\bibitem [{\citenamefont {Asakawa}\ and\ \citenamefont
  {Yazaki}(1989)}]{Asakawa:1989bq}%
  \BibitemOpen
  \bibfield  {author} {\bibinfo {author} {\bibfnamefont {M.}~\bibnamefont
  {Asakawa}}\ and\ \bibinfo {author} {\bibfnamefont {K.}~\bibnamefont
  {Yazaki}},\ }\href {\doibase 10.1016/0375-9474(89)90002-X} {\bibfield
  {journal} {\bibinfo  {journal} {Nucl. Phys. A}\ }\textbf {\bibinfo {volume}
  {504}},\ \bibinfo {pages} {668} (\bibinfo {year} {1989})}\BibitemShut
  {NoStop}%
\bibitem [{\citenamefont {Halasz}\ \emph {et~al.}(1998)\citenamefont {Halasz},
  \citenamefont {Jackson}, \citenamefont {Shrock}, \citenamefont {Stephanov},\
  and\ \citenamefont {Verbaarschot}}]{Halasz:1998qr}%
  \BibitemOpen
  \bibfield  {author} {\bibinfo {author} {\bibfnamefont {A.~M.}\ \bibnamefont
  {Halasz}}, \bibinfo {author} {\bibfnamefont {A.}~\bibnamefont {Jackson}},
  \bibinfo {author} {\bibfnamefont {R.}~\bibnamefont {Shrock}}, \bibinfo
  {author} {\bibfnamefont {M.~A.}\ \bibnamefont {Stephanov}}, \ and\ \bibinfo
  {author} {\bibfnamefont {J.}~\bibnamefont {Verbaarschot}},\ }\href {\doibase
  10.1103/PhysRevD.58.096007} {\bibfield  {journal} {\bibinfo  {journal} {Phys.
  Rev. D}\ }\textbf {\bibinfo {volume} {58}},\ \bibinfo {pages} {096007}
  (\bibinfo {year} {1998})},\ \Eprint {http://arxiv.org/abs/hep-ph/9804290}
  {arXiv:hep-ph/9804290} \BibitemShut {NoStop}%
\bibitem [{\citenamefont {Berges}\ and\ \citenamefont
  {Rajagopal}(1999)}]{Berges:1998rc}%
  \BibitemOpen
  \bibfield  {author} {\bibinfo {author} {\bibfnamefont {J.}~\bibnamefont
  {Berges}}\ and\ \bibinfo {author} {\bibfnamefont {K.}~\bibnamefont
  {Rajagopal}},\ }\href {\doibase 10.1016/S0550-3213(98)00620-8} {\bibfield
  {journal} {\bibinfo  {journal} {Nucl. Phys. B}\ }\textbf {\bibinfo {volume}
  {538}},\ \bibinfo {pages} {215} (\bibinfo {year} {1999})},\ \Eprint
  {http://arxiv.org/abs/hep-ph/9804233} {arXiv:hep-ph/9804233} \BibitemShut
  {NoStop}%
\bibitem [{\citenamefont {Skokov}\ \emph {et~al.}(2010)\citenamefont {Skokov},
  \citenamefont {Stokic}, \citenamefont {Friman},\ and\ \citenamefont
  {Redlich}}]{Skokov:2010wb}%
  \BibitemOpen
  \bibfield  {author} {\bibinfo {author} {\bibfnamefont {V.}~\bibnamefont
  {Skokov}}, \bibinfo {author} {\bibfnamefont {B.}~\bibnamefont {Stokic}},
  \bibinfo {author} {\bibfnamefont {B.}~\bibnamefont {Friman}}, \ and\ \bibinfo
  {author} {\bibfnamefont {K.}~\bibnamefont {Redlich}},\ }\href {\doibase
  10.1103/PhysRevC.82.015206} {\bibfield  {journal} {\bibinfo  {journal} {Phys.
  Rev.}\ }\textbf {\bibinfo {volume} {C82}},\ \bibinfo {pages} {015206}
  (\bibinfo {year} {2010})},\ \Eprint {http://arxiv.org/abs/1004.2665}
  {arXiv:1004.2665 [hep-ph]} \BibitemShut {NoStop}%
\bibitem [{\citenamefont {Skokov}\ \emph {et~al.}(2011)\citenamefont {Skokov},
  \citenamefont {Friman},\ and\ \citenamefont {Redlich}}]{Skokov:2010uh}%
  \BibitemOpen
  \bibfield  {author} {\bibinfo {author} {\bibfnamefont {V.}~\bibnamefont
  {Skokov}}, \bibinfo {author} {\bibfnamefont {B.}~\bibnamefont {Friman}}, \
  and\ \bibinfo {author} {\bibfnamefont {K.}~\bibnamefont {Redlich}},\ }\href
  {\doibase 10.1103/PhysRevC.83.054904} {\bibfield  {journal} {\bibinfo
  {journal} {Phys. Rev. C}\ }\textbf {\bibinfo {volume} {83}},\ \bibinfo
  {pages} {054904} (\bibinfo {year} {2011})},\ \Eprint
  {http://arxiv.org/abs/1008.4570} {arXiv:1008.4570 [hep-ph]} \BibitemShut
  {NoStop}%
\bibitem [{\citenamefont {Schaefer}\ and\ \citenamefont
  {Wambach}(2007)}]{Schaefer:2006ds}%
  \BibitemOpen
  \bibfield  {author} {\bibinfo {author} {\bibfnamefont {B.-J.}\ \bibnamefont
  {Schaefer}}\ and\ \bibinfo {author} {\bibfnamefont {J.}~\bibnamefont
  {Wambach}},\ }\href {\doibase 10.1103/PhysRevD.75.085015} {\bibfield
  {journal} {\bibinfo  {journal} {Phys. Rev. D}\ }\textbf {\bibinfo {volume}
  {75}},\ \bibinfo {pages} {085015} (\bibinfo {year} {2007})},\ \Eprint
  {http://arxiv.org/abs/hep-ph/0603256} {arXiv:hep-ph/0603256} \BibitemShut
  {NoStop}%
\bibitem [{\citenamefont {Almasi}\ \emph {et~al.}(2017)\citenamefont {Almasi},
  \citenamefont {Friman},\ and\ \citenamefont {Redlich}}]{Almasi:2017bhq}%
  \BibitemOpen
  \bibfield  {author} {\bibinfo {author} {\bibfnamefont {G.~A.}\ \bibnamefont
  {Almasi}}, \bibinfo {author} {\bibfnamefont {B.}~\bibnamefont {Friman}}, \
  and\ \bibinfo {author} {\bibfnamefont {K.}~\bibnamefont {Redlich}},\ }\href
  {\doibase 10.1103/PhysRevD.96.014027} {\bibfield  {journal} {\bibinfo
  {journal} {Phys. Rev. D}\ }\textbf {\bibinfo {volume} {96}},\ \bibinfo
  {pages} {014027} (\bibinfo {year} {2017})},\ \Eprint
  {http://arxiv.org/abs/1703.05947} {arXiv:1703.05947 [hep-ph]} \BibitemShut
  {NoStop}%
\bibitem [{\citenamefont {Tripolt}\ \emph {et~al.}(2021)\citenamefont
  {Tripolt}, \citenamefont {Jung}, \citenamefont {von Smekal},\ and\
  \citenamefont {Wambach}}]{Tripolt:2021jtp}%
  \BibitemOpen
  \bibfield  {author} {\bibinfo {author} {\bibfnamefont {R.-A.}\ \bibnamefont
  {Tripolt}}, \bibinfo {author} {\bibfnamefont {C.}~\bibnamefont {Jung}},
  \bibinfo {author} {\bibfnamefont {L.}~\bibnamefont {von Smekal}}, \ and\
  \bibinfo {author} {\bibfnamefont {J.}~\bibnamefont {Wambach}},\ }\href
  {\doibase 10.1103/PhysRevD.104.054005} {\bibfield  {journal} {\bibinfo
  {journal} {Phys. Rev. D}\ }\textbf {\bibinfo {volume} {104}},\ \bibinfo
  {pages} {054005} (\bibinfo {year} {2021})},\ \Eprint
  {http://arxiv.org/abs/2105.00861} {arXiv:2105.00861 [hep-ph]} \BibitemShut
  {NoStop}%
\end{thebibliography}%

\end{document}